\documentclass[twocolumn,aps,superscriptaddress,showpacs,nofootinbib,floatfix]{revtex4}
\usepackage{epsfig,bm,feynmf}
\usepackage{graphics}
\usepackage{amsmath}
\usepackage[normalem]{ulem}  
\usepackage[dvips]{color} 

\renewcommand\sout{\bgroup \color{red} \ULdepth=-.5ex \ULset}
\begin{document}


\title{Exploring non-equilibrium quark-gluon plasma effects on charm transport coefficients}


\author{Taesoo Song}\email{t.song@gsi.de}
\affiliation{GSI
Helmholtzzentrum f\"{u}r Schwerionenforschung GmbH, Planckstrasse 1,
64291 Darmstadt, Germany}

\author{Pierre Moreau}\email{pierre.moreau@duke.edu}
\affiliation{Institute for Theoretical Physics, Johann Wolfgang
Goethe Universit\"{a}t, Frankfurt am Main, Germany}
\affiliation{Department of Physics, Duke University, Durham, North Carolina 27708, USA}


\author{Joerg Aichelin}\email{aichelin@subatech.in2p3.fr}
\affiliation{SUBATECH UMR 6457 (IMT Atlantique, Universit\'{e} de Nantes,
	IN2P3/CNRS), 4 Rue Alfred Kastler, F-44307 Nantes, France}
\affiliation{FIAS, Ruth-Moufang-Straße 1, 60438 Frankfurt am Main, Germany}

\author{Elena Bratkovskaya}\email{brat@fias.uni-frankfurt.de}
\affiliation{GSI
Helmholtzzentrum f\"{u}r Schwerionenforschung GmbH, Planckstrasse 1,
64291 Darmstadt, Germany}
\affiliation{Institute for Theoretical Physics, Johann Wolfgang
Goethe Universit\"{a}t, Frankfurt am Main, Germany}


\begin{abstract}
We investigate how the drag coefficient $A$ and $\hat{q}$, the transverse momentum transfer by unit length, of charm quarks are modified if the QGP is not in complete thermal equilibrium using the dynamical quasi-particle model (DQPM) which reproduces both, the equation-of-state of the QGP and the spatial diffusion coefficient of heavy quarks as predicted by lattice quantum chromodynamics (lQCD) calculations. We study three cases:
a) the QGP has an anisotropic momentum distribution of the partons which leads to an anisotropic pressure
b) the QGP partons have higher or lower kinetic energies as compared to the thermal expectation value, and
c) the QGP partons have larger or smaller pole masses of their spectral function as compared to the pole mass from the DQPM at the QGP temperature.
In the last two cases we adjust the number density of partons to obtain the same energy density as in an equilibrated QGP.
We find that each non-equilibrium scenario affects $A$ and $\hat{q}$ of charm quarks in a different way. The modifications in our scenarios are of the order 20-30 \% at temperatures relevant for heavy ion reactions. These modifications have to be considered if one wants to determine these coefficients by comparing  heavy ion data with theoretical predictions from viscous hydrodynamics or Langevin equations.

\end{abstract}

\pacs{25.75.Nq, 25.75.Ld}
\keywords{}

\maketitle

\section{Introduction}

It is by now consesus that relativistic heavy-ion collisions produce an extremely hot and dense matter which transform into a QGP.
It is very important and valuable to investigate the properties of  matter under such extreme conditions, because such matter existed in  the early universe and  may exist in neutron stars. Heavy flavor particles are promising probes to study  matter at extreme conditions produced in heavy-ion collisions at relativistic energies. Their production is reliably described by perturbative quantum chromodynamics (pQCD). Since they are early produced,  in a  relativistic heavy-ion collision they collide with the bulk matter from the very early stage to the final one and collect information on these interactions.

The hot and dense matter modifies not only the production but also the dynamics of heavy flavor particles.
For example, heavy flavor particles with a  large transverse momentum, $p_\perp$, are pushed to lower  $p_\perp$  due to their energy-momentum loss in matter, while those with a small   $p_\perp$ are boosted toward larger  $p_\perp$ by the collective flow.
The interactions of heavy flavors with matter are expressed in terms of the nuclear modification factor, $R_{\rm AA}$, and the elliptic flow $v_2$.
It has been found that both $R_{\rm AA}$ at large $p_\perp$ and the elliptic flow, $v_2$, of heavy flavor particles are compatible to those observed for light flavors. This confirms that heavy flavor particles interact  strongly with matter~\cite{Sirunyan:2017xss,Acharya:2018hre,Adam:2018inb}

Since heavy flavor particles have much larger masses than their  light flavor counterparts, they are unlikely to thermalize in heavy-ion collisions unless they have initially a low  $p_\perp$. Therefore hydrodynamics is not applicable to heavy flavor particles, though it has been very successful for describing the dynamics of bulk particles.
Usually a Langevin or a Boltzmann transport approach is used to describe the time evolution of heavy flavor particles in heavy-ion collisions~\cite{Moore:2004tg,vanHees:2005wb,He:2014cla,Cao:2013ita,Gossiaux:2010yx,Das:2015ana,Song:2015sfa,Song:2015ykw,Rapp:2018qla,Cao:2018ews}.
The key ingredients of the Langevin approach as well as for viscous hydro calculations are transport coefficients which can be calculated from the scattering amplitudes of heavy flavor particle with light particles from the QGP~\cite{Svetitsky:1987gq,Moore:2004tg}. Usually the transport coefficients are calculated as functions of momentum and temperature using the assumption that the collision partners from the QGP are in thermal equilibrium.

For three reasons this assumption is probably not always true for heavy ion collisions to which these models are applied. 1) Even bulk particles need time to reach complete thermal equilibrium after production.
The color glass condensate model~\cite{Iancu:2000hn,Ferreiro:2001qy} predicts for example that  in the early stage of heavy-ion collisions the transverse pressure is stronger than the  longitudinal pressure. 2) The transverse momentum spectra of particles produced in heavy-ion collisions fall off with a power law for large momenta, a consequence of the hard scattering at production and are therefore far from a thermal distribution~\cite{Tsallis:1987eu,Song:2012at}. 3) Even in central collisions  there is a corona of nucleons which just do not scatter sufficiently often to come to a thermal equilibrium. This modifies transport coefficients as has been shown in ref. \cite{Aichelin:2010ns,Aichelin:2010ed}.

When comparing the calculated transport coefficients with experimental heavy ion data, these non-equilibrium effects have to be considered and it is therefore necessary and useful to study systematically how the transport coefficients of heavy quarks will be modified, if the QGP is not completely equilibrated.

In this study we calculate the transport coefficients of charm quarks using the collision integral for elastic collisions.  To describe the QGP we employ the dynamical quasiparticle model (DQPM)~\cite{Cassing:2009vt}, where a QGP is composed of strongly interacting quasiparticles whose pole mass and spectral width are fitted to the equation-of-state obtained by  lattice quantum chromodynamics (lQCD) calculation. The scattering between heavy quarks and the off-shell partons is calculated using leading order Feynman diagrams~\cite{Berrehrah:2014kba,Song:2016rzw}. This is motivated by the fact that the DQPM reproduces the spatial diffusion coefficient of heavy quarks as a function of temperature, which is presently the only observable for heavy quarks from lQCD.

We introduce three non-equilibrium scenarios in this study: a QGP with anisotropic pressure, a QGP with partons having excessive or deficient kinetic energies with respect to thermal ones, and a QGP where partons are heavier or lighter than the spectral function given by the DQPM. In the latter two cases we keep the equilibrium energy density by adjusting the particle density.
In each scenario the coefficients of momentum drag and transverse momentum diffusion of the charm quarks are calculated and compared to the coefficients obtained for a completely thermalized QGP.
In the non-equilibrium scenarios we calculate the temperature from the energy density with the help of the equation-of-state of partonic matter from lQCD.

This paper is organized as follows:
In section~\ref{DQPM} we briefly review the DQPM and the calculations of transport coefficients.
Then three non-equilibrium scenarios are introduced in section~\ref{3scenarios} and the transport coefficients of charm quark are calculated for each scenario and compared to those in equilibrium.
Finally all results are summarized in section~\ref{summary}, and transport coefficients in a simple case are derived in Appendix~\ref{derivations}.

\section{transport coefficients in DQPM}\label{DQPM}

\subsection{Dynamical Quasiparticle Model}

The dynamical quasiparticle model (DQPM) interprets the QGP in terms of strongly interacting quarks and gluons, whose properties are fitted to lattice QCD calculations in thermal equilibrium and at vanishing chemical potential. The self energies are complex. The real part gives the mean-field potentials, whereas the imaginary part provides information about the lifetime and/or reaction rates of the particles. The DQPM postulates the following form for the parton spectral function \cite{Linnyk:2015rco}:
\begin{align}
	\rho_{j}(\omega,{\bf p}) & = \frac{\gamma_{j}}{\tilde{E}_j}
	\left(\frac{1}{(\omega-\tilde{E}_j)^2+\gamma^{2}_{j}}
	-\frac{1}{(\omega+\tilde{E}_j)^2+\gamma^{2}_{j}}\right) \nonumber \\
	& \equiv \frac{4\omega\gamma_j}{\left( \omega^2 - \mathbf{p}^2 - M^2_j \right)^2 + 4\gamma^2_j \omega^2}
	\label{spectral_function}
\end{align}
~\\
separately for quarks, antiquarks, and gluons ($j = q,\bar q,g$).
Here, $\tilde{E}_{j}^2({\bf p})={\bf p}^2+M_{j}^{2}-\gamma_{j}^{2}$,  the widths $\gamma_{j}$ and the masses $M_{j}$ from the DQPM are functions of the temperature $T$ and the chemical potential $\mu$. Both are chosen to reproduce the lQCD calculations.
By applying a nonrelativistic approximation to Eq. (\ref{spectral_function}), one obtains Breit-Wigner spectral functions, $\rho(m)$, which read as \cite{Berrehrah:2013mua}
\begin{equation}
\rho_i(m) = \frac{2}{\pi} \frac{2m^2 \gamma_i}{(m^2-M_i^2)^2+(2m \gamma_i)^2}.
\label{Breit-Wigner}
\end{equation}
~\\
This distribution fulfills the normalization $\int_{0}^{\infty} dm\ \rho(m) = 1$.

The dynamical quasiparticle mass $M$ is assumed to be given by the HTL thermal mass in the asymptotic high-momentum regime, i.e., for gluons by \cite{Linnyk:2015rco}

\begin{equation}
M^2_{g}(T,\mu_q)=\frac{g^2(T,\mu_q)}{6}\left(\left(N_{c}+\frac{1}{2}N_{f}\right)T^2
+\frac{N_c}{2}\sum_{q}\frac{\mu^{2}_{q}}{\pi^2}\right)\ ,
\label{Mg9}
\end{equation}
~\\
and for quarks (antiquarks) by

\begin{equation}
M^2_{q(\bar q)}(T,\mu_q)=\frac{N^{2}_{c}-1}{8N_{c}}g^2(T,\mu_q)\left(T^2+\frac{\mu^{2}_{q}}{\pi^2}\right)\ ,
\label{Mq9}
\end{equation}
~\\
where $N_{c}=3$ stands for the number of colors while $N_{f}\ (=3)$ denotes the number of flavors. Furthermore, the effective quarks, antiquarks, and gluons in the DQPM have finite widths $\gamma$, which are adopted in the form \cite{Linnyk:2015rco}

\begin{equation}
\label{widthg}
\gamma_{g}(T,\mu_q) = \frac{1}{3}N_{c}\frac{g^2(T,\mu_q)T}{8\pi}\ln\left(\frac{2c}{g^2(T,\mu_q)}+1\right),
\end{equation}
\begin{equation}
\label{widthq}
\gamma_{q(\bar
	q)}(T,\mu_q)=\frac{1}{3}\frac{N^{2}_{c}-1}{2N_{c}}\frac{g^2(T,\mu_q)T}{8\pi}
\ln\left(\frac{2c}{g^2(T,\mu_q)}+1\right),
\end{equation}
~\\
where $c=14.4$  is related to a magnetic cutoff, which is an additional parameter of the DQPM. Furthermore, we assume that the width of the strange quarks is the same as that for the light ($u,d$) quarks. With the choice of Eq. (\ref{spectral_function}), the complex self-energies for gluons $\Pi = M_g^2-2i \omega \gamma_g$ and for (anti)quarks $\Sigma_{q} = M_{q}^2 - 2 i \omega \gamma_{q}$ are fully defined via Eqs. (\ref{Mg9}), (\ref{Mq9}), (\ref{widthg}), and (\ref{widthq}).

The coupling $g^2$, which defines the strength of the interaction in the DQPM, is extracted from lQCD thermodynamics. Its temperature dependence at vanishing chemical potential can either be obtained by using an ansatz with a few parameters  adjusted to results of lQCD thermodynamics \cite{Berrehrah:2013mua,Berrehrah:2014ysa,Berrehrah:2015ywa}, or can directly be obtained by a parametrization of the entropy density from lQCD as explained in Ref. \cite{Moreau:2019vhw}. The extension to finite baryon chemical potential, $\mu_B$, is performed by using a scaling ansatz which works up to $\mu_B \approx 450$ MeV \cite{Steinert:2018bma}, and which assumes that $g^2$ is a function of the ratio of the effective temperature $T^* = \sqrt{T^2+\mu^2_q/\pi^2}$ and the $\mu_B$-dependent critical temperature $T_c(\mu_B)$ as \cite{Cassing:2007nb}

\begin{equation}
g^2(T/T_c,\mu_B) = g^2\left(\frac{T^*}{T_c(\mu_B)},\mu_B =0 \right)
\label{coupling}
\end{equation}
~\\
\noindent with $\mu_B=3\mu_q$ and $T_c(\mu_B) = T_c \sqrt{1-\alpha \mu_B^2}$, where $T_c$ is
the critical temperature at vanishing chemical potential ($\approx 0.158$ GeV) and $\alpha= 0.974\ \text{GeV}^{-2}$.

By employing the quasiparticle properties and dressed propagators as given by the DQPM, one can deduce the scattering cross sections as well as the interaction rates of light and charm quarks in the QGP as a function of the temperature and the chemical potential \cite{Berrehrah:2013mua,Moreau:2019vhw}. The matrix elements $|M|^2$ of the corresponding processes are calculated in leading order.
For simplicity, we restrict ourselves  in this study to vanishing baryon chemical potential, which reflects the physical situation in heavy-ion collisions at the relativistic heavy-ion collider (RHIC) and the large hadron collider (LHC).

\subsection{transport coefficients}

The distribution of charm quarks changes in a matter with time due to interactions and is expressed in a Fokker-Planck equation as

\begin{eqnarray}
\frac{\partial f(p)}{\partial t}=\frac{\partial}{\partial p_i}\bigg[A_i(p)f(p) +\frac{\partial}{\partial p_i}[B_{ij}(p)f(p)]\bigg],
\end{eqnarray}
where $f(p)$ is charm distribution function~\cite{Svetitsky:1987gq}.
Here, we study the drag and transverse/longitudinal diffusion coefficients which are defined as~\cite{Svetitsky:1987gq,Moore:2004tg,Berrehrah:2014kba}

\begin{eqnarray}
A(p)&=&-\frac{\vec{A}(p)\cdot \vec{p}}{|\vec{p}|}= -\frac{d\langle \Delta p \rangle}{dt}=\langle\langle(p-p^\prime)_x\rangle\rangle,\label{A}\\
B_L(p)&=& \frac{1}{2}\frac{p_ip_j}{|\vec{p}|^2}B_{ij}(p)\nonumber\\
&=&\frac{1}{2}\frac{d\langle (\Delta p_L)^2 \rangle}{dt}= \frac{1}{2}\langle\langle(p-p^\prime)_x^2\rangle\rangle,\label{bldef}\\
B_T(p)&=& \frac{1}{4}\bigg(\delta_{ij}-\frac{p_ip_j}{|\vec{p}|^2}\bigg)B_{ij}(p)\nonumber\\
&=&\frac{1}{4}\frac{d\langle (\Delta p_T)^2 \rangle}{dt}= \frac{1}{4}\langle\langle p_y^{\prime 2}+p_z^{\prime 2}\rangle\rangle,\label{btdef}\\
\hat{q}(p)&=&\frac{d\langle(\Delta p_T)^2\rangle}{d x}=\frac{4E}{p_L}B_T(p),
\label{drag}
\end{eqnarray}
where
\begin{eqnarray}
\langle\langle O^* \rangle\rangle \equiv \frac{1}{2E_p}\sum_{i=q,\bar{q},g}\int \frac{d^3k}{(2\pi)^3 2E}f_i(k)\int \frac{d^3k^\prime}{(2\pi)^3 2E^\prime}\nonumber\\
\times\int \frac{d^3p^\prime}{(2\pi)^3 2E_p^\prime}~O^*~ (2\pi)^4\delta^{(4)}(p+k-p^\prime-k^\prime)\frac{|M_{ic}|^2}{\gamma_c},
\label{def1}
\end{eqnarray}
with $(E_p,~p),~(E,~k)$ being energy-momenta in the entrance channel of the heavy flavor parton and of the scattering partner $i$ from the QGP, respectively.  $(E_p^\prime, ~p^\prime),~(E^\prime, ~k^\prime)$  are the  energy-momenta after scattering and $M_{ic}$, $\gamma_c$, and $f_i(k)$ are the scattering amplitude, the degeneracy factor of the particle of interest, and the distribution function of scattering partner $i$, respectively.
Since we are interested in heavy flavor particles at mid-rapidity ($p_z=0$), the initial momentum is taken to be $\vec{p}=(p_x,0,0)$ in the above equations.
We note that the definitions of $B_L(p)$ and $B_T(p)$ from Ref.~\cite{Moore:2004tg} are a factor of two larger than those from Ref.~\cite{Svetitsky:1987gq} which we adopt in Eqs.~(\ref{bldef}) and (\ref{btdef}).

Equation~(\ref{def1}) is equivalent to
\begin{eqnarray}
\langle\langle O^* \rangle\rangle \equiv \sum_{i=q,\bar{q},g}\int \frac{d^3k}{(2\pi)^3 }\gamma_i f_i(k)~O^*~ v_{ic}\sigma_{ic},
\label{def2}
\end{eqnarray}
where $v_{ic}$ and $\sigma_{ic}$ are the relative velocity and the scattering cross section of particle $c$ and $i$, respectively.

Since partons in the DQPM are off-shell particles, their mass has a spectral distribution, and Eqs.~(\ref{def1}) and (\ref{def2}) are modified into

\begin{eqnarray}
\langle\langle O^* \rangle\rangle \equiv \frac{1}{2E_p}\sum_{i=q,\bar{q},g}\int dm \rho_i(m)\int dm^\prime \rho_i(m^\prime)\nonumber\\
\times \int \frac{d^3k}{(2\pi)^3 2E}f_i(k,m_i)
\int \frac{d^3k^\prime}{(2\pi)^3 2E^\prime}\int \frac{d^3p^\prime}{(2\pi)^3 2E_p^\prime}\nonumber\\
\times O^*~ (2\pi)^4\delta^{(4)}(p+q-p^\prime-q^\prime)\frac{|M_{ic}|^2}{\gamma_c},
\end{eqnarray}
or
\begin{eqnarray}
\langle\langle O^* \rangle\rangle \equiv \sum_{i=q,\bar{q},g}\int dm \rho_i(m)\int \frac{d^3k}{(2\pi)^3 }\gamma_i f_i(k,m_i)\nonumber\\
\times O^*~ v_{ic}\overline{\sigma_{ic}},
\label{drag2}
\end{eqnarray}
where $\rho_i(m)$ and $\rho_i(m^\prime)$ are the spectral functions of the incoming and the outgoing light partons, Eq.~(\ref{Breit-Wigner}). $\overline{\sigma_{ic}}$ is the scattering cross section averaged over the final mass of the light parton.
We assume that the off-shellness of charm quarks is negligible. This means that charm quarks have a fixed mass.

Since the mass of the incoming parton is different from that of the outgoing parton, the integration over the phase space has to be done carefully, especially when the outgoing mass is larger than the  incoming mass. If the energy-momenta of the charm quark and the light parton are expressed as

\begin{eqnarray}
p^\mu&=&(E_p,~p,~0,~0),\nonumber\\
k^\mu&=&(E,~k\cos\theta,~k\sin\theta,~0),\nonumber
\end{eqnarray}
the threshold energy for the production of a more massive final state is given by

\begin{eqnarray}
s=m_c^2+m_i^2+2E_pE-2pk\cos\theta\geq(m_c+m_f)^2,
\end{eqnarray}
where $m_c^2=p^2$, $m_i^2=k^2$ and $m_f^2=k^{\prime 2}$. This leads to
\begin{eqnarray}
\frac{m_i^2-m_f^2+2(E_pE-m_cm_f)}{2pk}\geq \cos\theta.
\label{condition}
\end{eqnarray}
If the left-hand-side of Eq.~(\ref{condition}) equals $-1$,  $\theta$ = $\pi$ and the corresponding momentum is given by
\begin{eqnarray}
k=\frac{-Cp+E_p\sqrt{C^2-m_i^2m_c^2}}{m_c^2},
\label{11}
\end{eqnarray}
where $C\equiv m_cm_f+(m_f^2-m_i^2)/2$.
Assuming $m_f>m_i$, $C$ is always positive.

On the other hand, if the left-hand-side of Eq.~(\ref{condition}) equals $+1$, all values of $\theta$ are possible and the corresponding momentum is
\begin{eqnarray}
k=\frac{Cp+E_p\sqrt{C^2-m_i^2m_c^2}}{m_c^2},
\label{12}
\end{eqnarray}
or
\begin{eqnarray}
k=\frac{Cp-E_p\sqrt{C^2-m_i^2m_c^2}}{m_c^2}.
\label{13}
\end{eqnarray}

\begin{figure} [h!]
\centerline{
\includegraphics[width=8. cm]{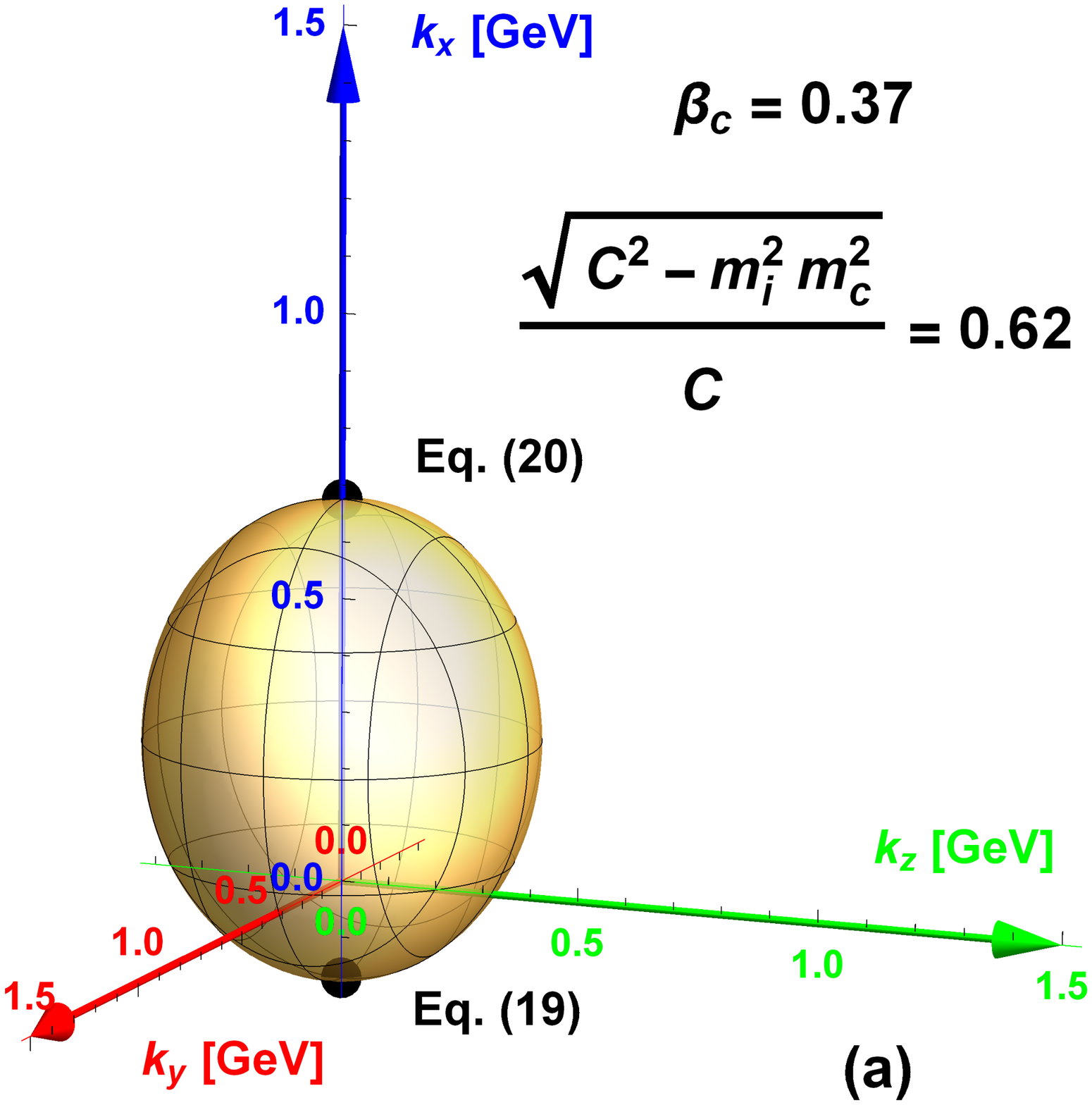}}
\vspace{10mm}
\centerline{
\includegraphics[width=8. cm]{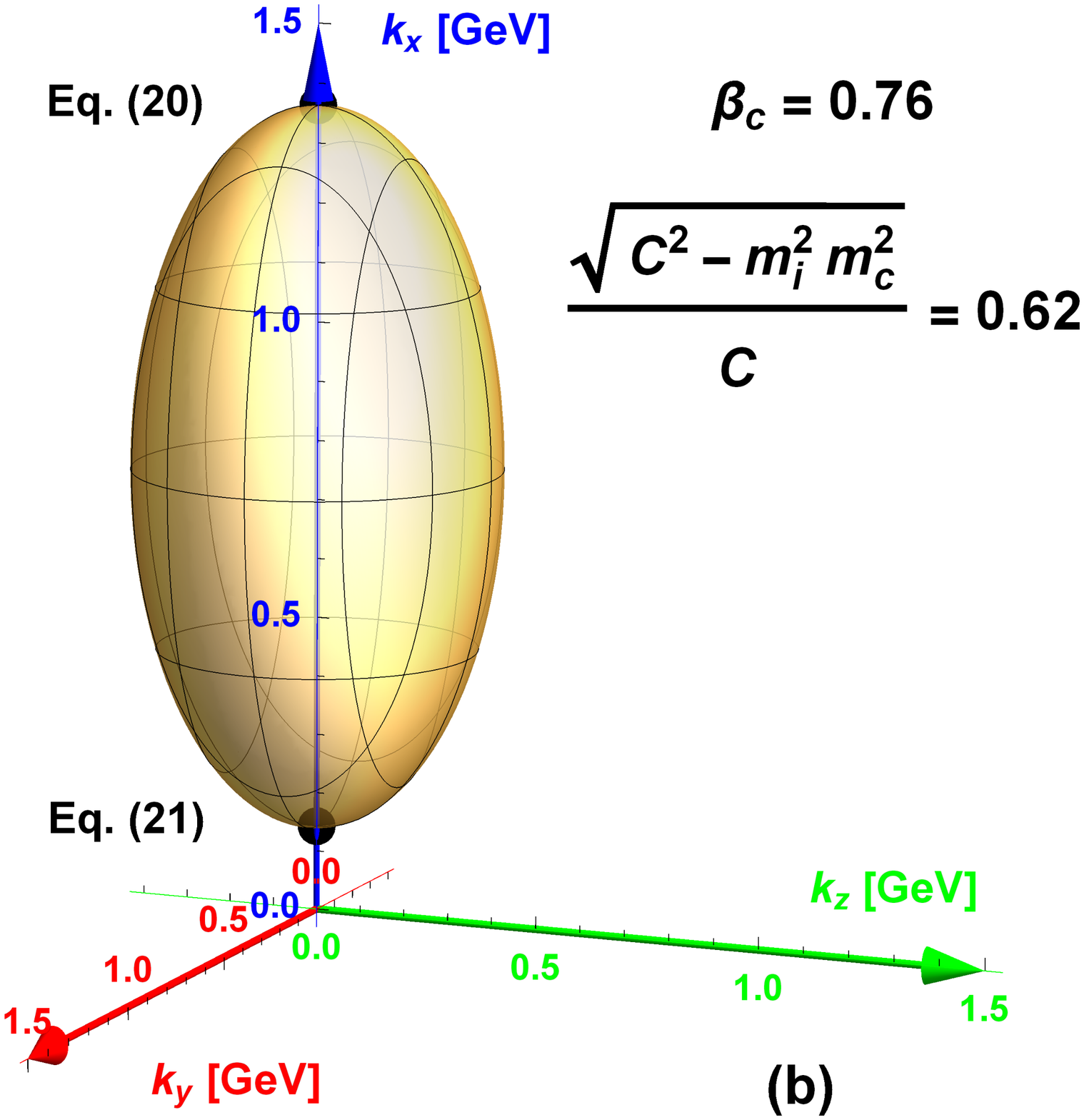}}
\caption{$\vec{k}$ is kinematically forbidden inside the volumes (a) for a small charm quark momentum $p$ and (b) for large $p$.
Two intersections of the $k_x$ line, which indicates the direction of $\vec{p}$, by the 3-dimensional volume correspond to Eqs.~(\ref{11}) and (\ref{12}) in the upper figure and to Eqs.~(\ref{12}) and (\ref{13}) in the lower figure.}
\label{kinematic}
\end{figure}

The kinematically forbidden range of $\vec{k}$ is depicted in figure~\ref{kinematic}.
The inside of the volume in figure~\ref{kinematic} (a) indicates the kinematically forbidden $\vec{k}$ for small charm momentum and the volume in figure~\ref{kinematic} (b) for large charm momentum.
The $k_x$ line indicates the direction of $\vec{p}$.
In the former case the maximum and the minimum of forbidden $k$ are given by Eqs.~(\ref{12}) and (\ref{11}),  respectively,  in the latter case by  Eqs.~(\ref{12}) and (\ref{13}).
The two cases coincide, when

\begin{eqnarray}
E_p=E_p^*\equiv\frac{m_f^2-m_i^2+2m_cm_f}{2m_i}=\frac{C}{m_i}.
\label{coincide}
\end{eqnarray}

In other words, if $E_p$ is smaller than Eq.~(\ref{coincide}), it corresponds to the upper figure, and larger $E_p$ corresponds to the lower figure.
The integration range of $\cos\theta$ is given by Eq.~(\ref{condition}) in both cases.

\begin{figure} [h!]
\centerline{
\includegraphics[width=8.6 cm]{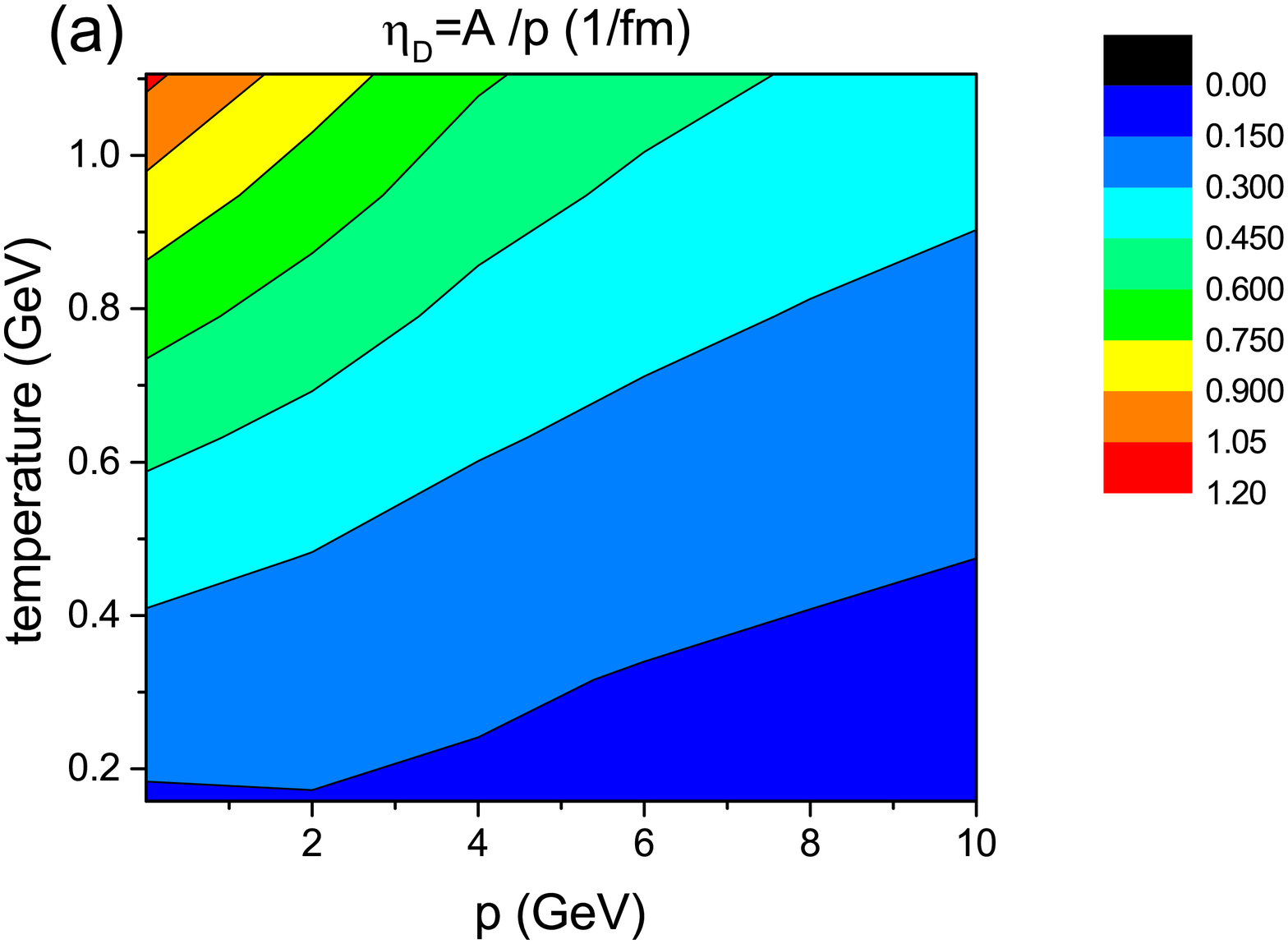}}
\centerline{
\includegraphics[width=8.6 cm]{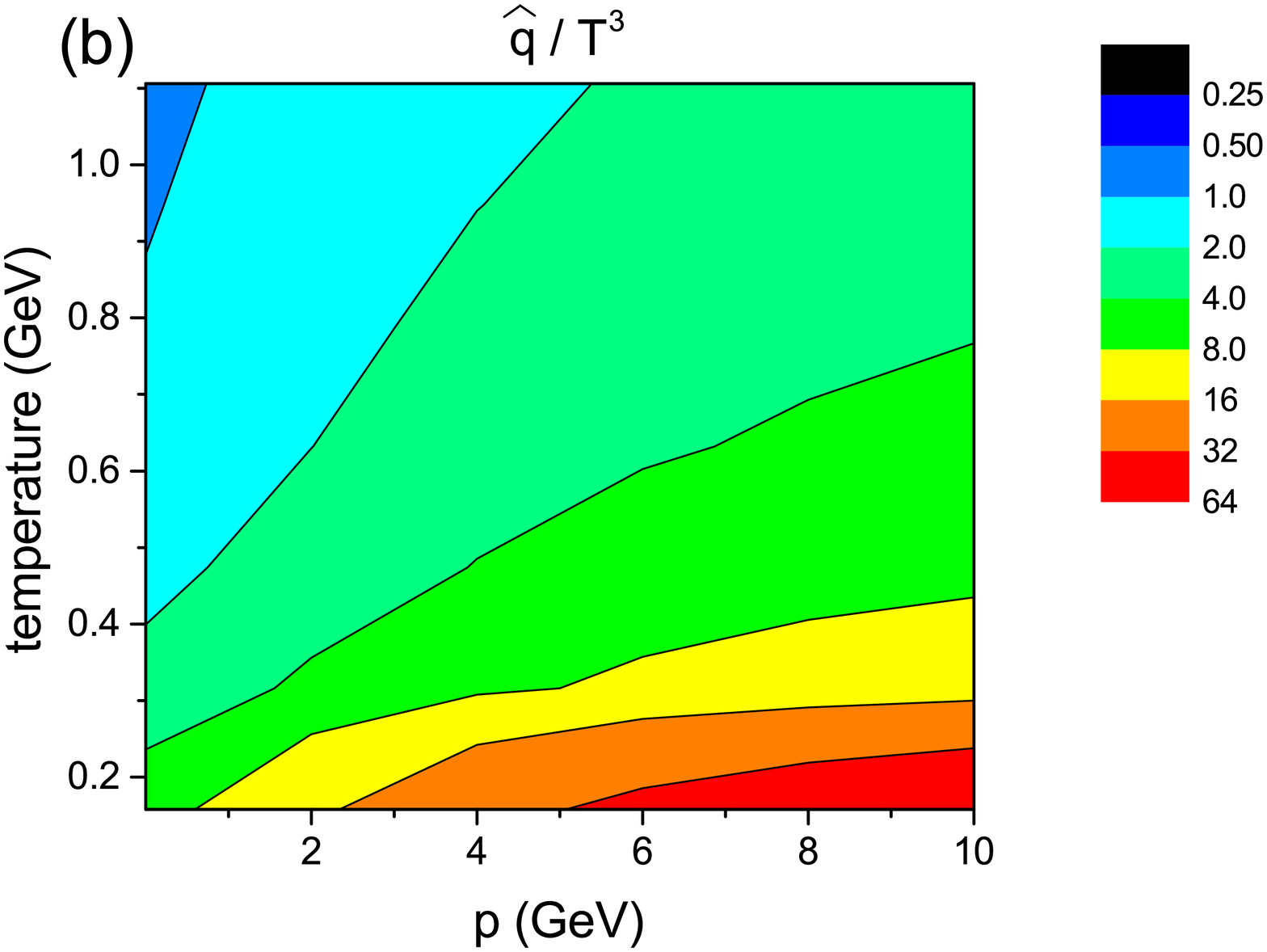}}
\centerline{
\includegraphics[width=8.6 cm]{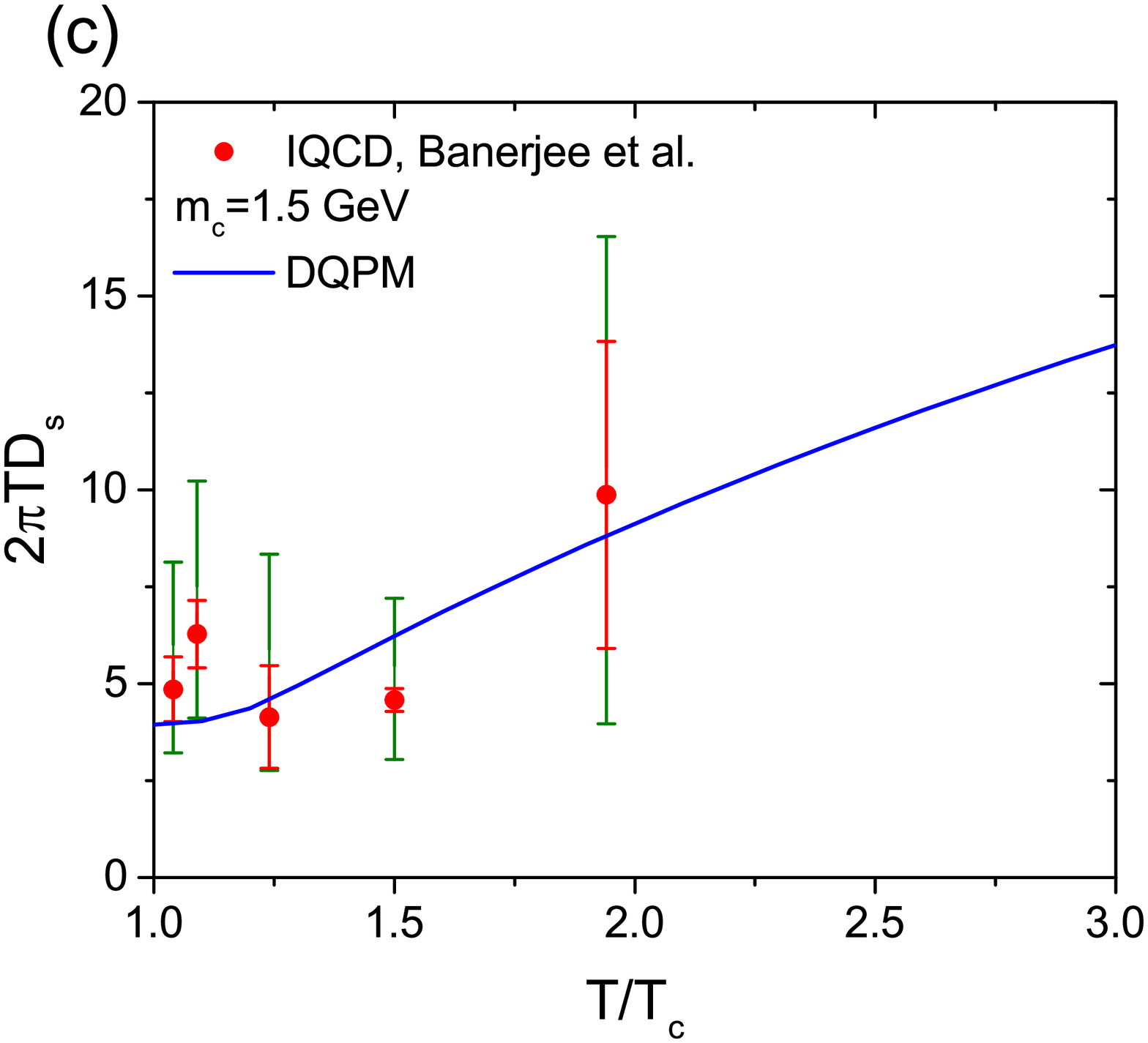}}
\caption{Drag coefficient divided by the charm quark momentum (upper), $\hat{q}$ scaled by $T^3$ (middle), and the spatial diffusion coefficient (lower) of charm quarks as a function of temperature and/or charm quark momentum from the DQPM~\cite{Berrehrah:2014kba,Song:2016rzw} in comparison with the spatial diffusion coefficients from lQCD calculations~\cite{Banerjee:2011ra}.} \label{Ds}
\end{figure}

We show the drag coefficient divided by charm quark momentum in the upper panel of figure~\ref{Ds} and $\hat{q}=d<p_t^2>/v_Ldt$ of charm quarks (with $v_L$ being the longitudinal velocity) scaled by $T^3$ in the middle panel
as functions of temperature and charm quark momentum. The lower figure presents the spatial diffusion coefficients as a function of temperature, which corresponds to the vertical axis of the upper panel:

\begin{eqnarray}
D_s=\lim_{p\rightarrow 0}\frac{T}{m_c \eta},
\end{eqnarray}
where $\eta=A/p$ with $p$ being charm quark momentum~\cite{Berrehrah:2014kba}.
The spatial diffusion coefficient is small near $T_c$ in spite of a small parton number density, because the strong coupling increases rapidly near $T_c$ in the DQPM~\cite{Berrehrah:2014kba}.
As follows from the figure, the spatial diffusion coefficients from the DQPM are comparable to those from lattice QCD (lQCD) calculations~\cite{Banerjee:2011ra}.

\section{modeling of non-equilibrium matter}\label{3scenarios}

\subsection{scenario 1: modeling of anisotropic medium}\label{aniso-pressure}

We start with the modeling of an anisotropic medium.
The longitudinal and transverse pressures with respect to the beam direction are respectively defined as
\begin{eqnarray}
P_\|=T_{zz}=\int \frac{d^3k}{(2\pi)^3}\frac{k_z^2}{E}f(k),\label{pressurel}\\
P_\perp=\frac{T_{xx}+T_{yy}}{2}=\int \frac{d^3k}{(2\pi)^3}\frac{k_x^2+k_y^2}{2E}f(k),
\label{pressuret}
\end{eqnarray}
where $f(k)$ is the particle distribution function and $z$ is the beam direction of the heavy-ion collision.
We note that the longitudinal and transverse directions in the above equations are different from those in Eqs.~(\ref{A})-(\ref{drag}) which are defined with respect to the charm quark direction.

In isotropic or thermalized matter $P_\|/P_\perp =1$ while $P_\|/P_\perp \neq 1$ for anisotropic matter.
For example, the glasma from the color glass condensate model has a smaller longitudinal pressure compared to the transverse pressure  $P_\|/P_\perp <1$~\cite{Gelis:2013rba,Chen:2015wia,Song:2015jmn} before the thermalization time.
On the other hand, in the parton-hadron-string dynamics (PHSD) transport approach the transverse pressure is initially small and grows with time before reaching thermal equilibrium~\cite{Xu:2017pna}.

An anisotropic pressure can be modeled in many different ways.
For instance, the magnitude of the momentum vector of the particle is rescaled as a function of the momentum angle or an  angle dependent weight function of momentum vector is introduced.
While the effect of the magnitude of the  momentum vector will be discussed in the next subsection, in this subsection here we generate an anisotropic pressure by introducing a weight function to $f(k)$ as follows:
For $P_\|/P_\perp >1$,

\begin{eqnarray}
f(k)\rightarrow f(k,\theta)=(\alpha+1) |\cos\theta|^\alpha f(k),~(\alpha\geq 0),
\label{cos}
\end{eqnarray}
where $\cos\theta=k_z/|{\bf k}|$. $\alpha$ is a parameter to control the anisotropy, and `$\alpha+1$' is the normalization factor to keep particle number per volume the same.

For $P_\|/P_\perp <1$,
\begin{eqnarray}
f(k)\rightarrow f(k,\theta)=(\alpha+1) (1-|\cos\theta|)^\alpha f(k),~(\alpha\leq 0).
\label{sin}
\end{eqnarray}
Eqs. (\ref{cos}) and (\ref{sin}) are combined into
\begin{eqnarray}
f(k)\rightarrow f(k,\theta)=(|\alpha|+1)~~~~~~~~~~\nonumber\\
\times \bigg\{|\cos\theta|^\alpha\Theta(\alpha)+(1-|\cos\theta|)^{-\alpha}\Theta(-\alpha)\bigg\}f(k),
\label{both}
\end{eqnarray}
where $\Theta(x)$ is the step function.
In other words, $P_\|/P_\perp >1$ for $\alpha>0$ and $P_\|/P_\perp <1$ for $\alpha<0$.
Integrating Eqs. (\ref{pressurel}) and (\ref{pressuret}) over $\cos\theta$ we obtain

\begin{eqnarray}
P_\|&\sim& \int_0^1 d(\cos\theta) \cos^{\alpha+2}\theta=\frac{1}{\alpha+3},\nonumber\\
P_\perp &\sim& \frac{1}{2}\int_0^1 d(\cos\theta) \sin^2\theta\cos^\alpha\theta=\frac{1}{(\alpha+1)(\alpha+3)},\nonumber\\
\label{pressure1}
\end{eqnarray}
for $\alpha \geq0$, and

\begin{eqnarray}
P_\|&\sim& \int_0^1 d(\cos\theta) \cos^2\theta (1-\cos\theta)^{-\alpha}\nonumber\\
&=&\frac{2}{(-\alpha+1)(-\alpha+2)(-\alpha+3)},\nonumber\\
P_\perp &\sim& \frac{1}{2}\int_0^1 d(\cos\theta) \sin^2\theta (1-\cos\theta)^{-\alpha}\nonumber\\
&=&\frac{4-\alpha}{2(-\alpha+2)(-\alpha+3)}
\label{pressure2}
\end{eqnarray}

for $\alpha \leq 0$.
Therefore the relationship between pressure anisotropy and $\alpha$ is given by:
\begin{eqnarray}
P_\|/P_\perp = \alpha+1,~~~(\alpha\geq 0),\nonumber\\
P_\|/P_\perp = \frac{4}{(-\alpha+1)(-\alpha+4)},~~~(\alpha\leq 0),
\label{anisotropy}
\end{eqnarray}
with $P_\|/P_\perp=1$ at $\alpha=0$.

\begin{figure} [h!]
\centerline{
\includegraphics[width=8.6 cm]{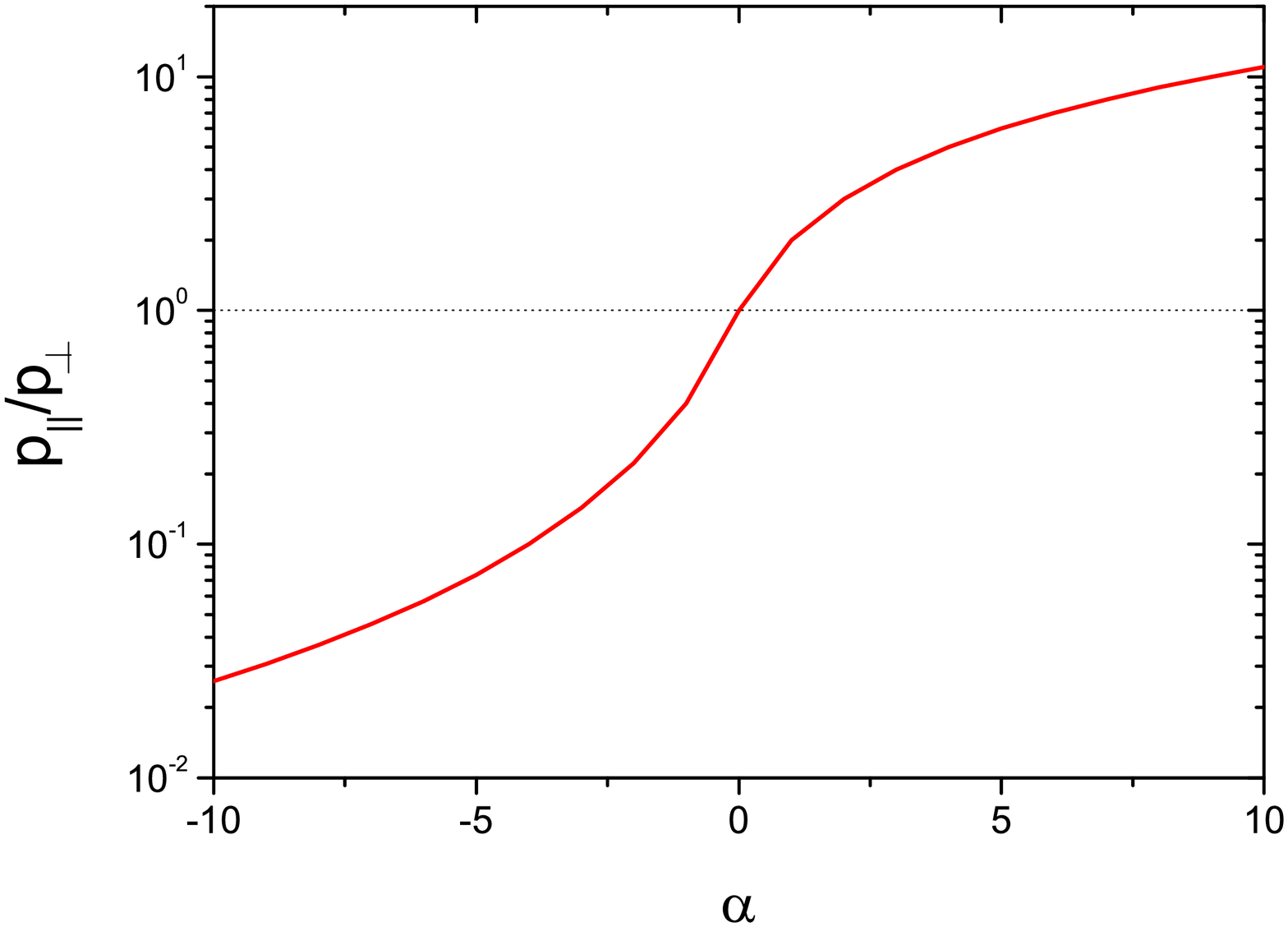}}
\caption{Pressure anisotropy as a function of $\alpha$, see Eq.~(\ref{anisotropy})} \label{alpha}
\end{figure}

Figure~\ref{alpha} shows the pressure anisotropy as a function of $\alpha$.
As mentioned, the pressure is isotropic for $\alpha=0$, the longitudinal pressure is larger than the transverse pressure for $\alpha >0$, and vice versa for $\alpha <0$.

To get an idea how the transport coefficients depend on the anisotropy of the pressure we consider some simple cases.
The interaction rate in the heat bath frame is given by
\begin{eqnarray}
\frac{dN^{\rm coll}}{dt}=\langle\langle 1 \rangle\rangle=\int \frac{d^3p_2}{(2\pi)^3}f(p_2)v_{\rm rel}\sigma~~~~~~~~~~~~~~~~~~\nonumber\\
=\frac{1}{2E_1}\int \frac{d^3p_2}{(2\pi)^32E_2}f(p_2)\int\frac{d^3p_3}{(2\pi)^32E_3}\int\frac{d^3p_4}{(2\pi)^32E_4}\nonumber\\
\times |\overline{M}|^2(2\pi)^4\delta^{(4)}(p_1+p_2-p_3-p_4),
\end{eqnarray}
where $\langle\langle ... \rangle\rangle$ is defined in Eq.~(\ref{def1}) and all particles are assumed to be on-shell.
Assuming for simplicity isotropic scattering with $|\overline{M}|^2=1$, we obtain 
\begin{eqnarray}
\frac{dN^{\rm coll}}{dt}=\frac{1}{8\pi E_1}\int \frac{d^3p_2}{(2\pi)^32E_2}f(p_2)\frac{p^{cm}}{\sqrt{s}},
\label{int-rate}
\end{eqnarray}
where
\begin{eqnarray}
p^{cm}=\frac{\sqrt{\{s-(m_3+m_4)^2\}\{s-(m_3-m_4)^2\}}}{2\sqrt{s}}\nonumber
\end{eqnarray}
with $m_3^2=p_3^2$ and $m_4^2=p_4^2$.

Now we move to the center-of-mass frame.
Since the scattering is isotropic, the average $\Delta p_L$, $(\Delta p_L)^2$, and $(\Delta p_T)^2$ per scattering are simply calculated as ~\cite{Berrehrah:2014kba}:
\begin{eqnarray}
\langle\Delta p_L^{cm}\rangle&=&\frac{p^{cm}}{4\pi}\int d\Omega(\cos\theta-1)=-p^{cm},\nonumber\\
\langle(\Delta p_L^{cm})^2\rangle&=&\frac{(p^{cm})^2}{4\pi}\int d\Omega(\cos\theta-1)^2=\frac{4}{3}(p^{cm})^2,\nonumber\\
\langle(\Delta p_T^{cm})^2\rangle&=&\frac{(p^{cm})^2}{4\pi}\int d\Omega\sin^2\theta=\frac{2}{3}(p^{cm})^2.
\label{dpt2}
\end{eqnarray}
The result can be expressed in vector or tensor form
\begin{eqnarray}
\langle\Delta p_i^{cm}\rangle&=&-p^{cm}(\widehat{p_1}^{cm})_i,\label{dpl}\\
\langle\Delta p_i^{cm}\Delta p_j^{cm}\rangle&=&\frac{1}{2}\{\delta_{ij}-(\widehat{p_1}^{cm})_i(\widehat{p_1}^{cm})_j\}\langle(\Delta p_T^{cm})^2\rangle\nonumber\\
&&+(\widehat{p_1}^{cm})_i(\widehat{p_1}^{cm})_j\langle(\Delta p_L^{cm})^2\rangle\nonumber\\
&=&\bigg\{\frac{\delta_{ij}}{3}+(\widehat{p_1}^{cm})_i(\widehat{p_1}^{cm})_j\bigg\}(p^{cm})^2,
\label{tensor}
\end{eqnarray}
where $\widehat{p_1}^{cm}$ is the unit vector of $\vec{p_1}$ in the center-of-mass frame.
The result can be brought into a covariant form, $\langle\Delta p_\mu^{cm}\rangle$ and $\langle\Delta p_\mu^{cm}\Delta p_\nu^{cm}\rangle$ with vanishing time components,
\begin{eqnarray}
\langle\Delta p_0^{cm}\rangle&=&0,\nonumber\\
\langle\Delta p_0^{cm}\Delta p_\mu^{cm}\rangle&=&\langle\Delta p_\mu^{cm}\Delta p_0^{cm}\rangle=0,
\label{end}
\end{eqnarray}
because the energy does not change by elastic scattering in the center-of-mass frame. In the DQPM model
energy can be exchanged even in elastic collisions, because the final mass may be different from the initial one.
This corresponds to a quasi-elastic scattering.

Moving back to heat bath frame and using Eqs.~(\ref{int-rate})-(\ref{end}), one can construct drag and diffusion coefficients as follows:
\begin{eqnarray}
&&A(p_1)= -\frac{p^{cm}}{8\pi E_1\sqrt{s}}(\widehat{p_1})_i\nonumber\\
&&~~~~\times \int \frac{d^3p_2}{(2\pi)^32E_2}f(p_2) \widetilde{L}_i^\mu \langle\Delta p_\mu^{cm}\rangle,\label{final-A}\\
&&B_L(p_1)=\frac{p^{cm}}{16\pi E_1\sqrt{s}}(\widehat{p_1})_i(\widehat{p_1})_j \nonumber\\
&&~~~~\times \int \frac{d^3p_2}{(2\pi)^32E_2}f(p_2) \widetilde{L}_{i}^{\mu}\widetilde{L}_{j}^{\nu} \langle\Delta p_\mu^{cm}\Delta p_\nu^{cm}\rangle,\\
&&B_T(p_1)=\frac{p^{cm}}{32\pi E_1\sqrt{s}}\bigg\{\delta_{ij}-(\widehat{p_1})_i(\widehat{p_1})_j\bigg\}\nonumber\\
&&~~~~\times \int \frac{d^3p_2}{(2\pi)^32E_2}f(p_2) \widetilde{L}_{i}^{\mu}\widetilde{L}_{j}^{\nu} \langle\Delta p_\mu^{cm}\Delta p_\nu^{cm}\rangle,\\
&&\hat{q}(p_1)=\frac{p^{cm}}{8\pi p_1\sqrt{s}}\bigg\{\delta_{ij}-(\widehat{p_1})_i(\widehat{p_1})_j\bigg\}\nonumber\\
&&~~~~\times \int \frac{d^3p_2}{(2\pi)^32E_2}f(p_2) \widetilde{L}_{i}^{\mu}\widetilde{L}_{j}^{\nu} \langle\Delta p_\mu^{cm}\Delta p_\nu^{cm}\rangle,
\label{general}
\end{eqnarray}
where $\widehat{p_1}$ is the unit vectors of $\vec{p_1}$ in the heat bath frame, and $\widetilde{L}$ symbolizes the Lorentz transformation from the center-of-mass frame to the heat bath frame.
We note that $\widetilde{L}_i^\mu (\widehat{p_1}^{cm})_\mu\neq (\widehat{p_1})_i$ because $\widehat{p_1}^{cm}$ has only spatial components.
Since we are interested in charm at mid-rapidity, $\widehat{p_1}$ is  in the $x-$direction, $\widehat{x}$.

For a simple check, we take the following three distribution functions:
\begin{eqnarray}
{\rm case~ 1:}~f(p_2)&=&\frac{(2\pi)^3}{2V}\delta(p_{2x})\delta(p_{2y})\nonumber\\
&&\times\bigg\{\delta(p_{2z}-k)+\delta(p_{2z}+k)\bigg\},\label{case1}\\
{\rm case~ 2:}~f(p_2)&=&\frac{(2\pi)^3}{2\pi V}\delta(p_{2x}-k\cos\phi)\nonumber\\
&&\times\delta(p_{2y}-k\sin\phi)\delta(p_{2z}),\label{case2}\\
{\rm case~ 3:}~f(p_2)&=&\frac{(2\pi)^3}{4\pi V}\delta(p_{2x}-k\cos\theta)\delta(p_{2y}-k\sin\theta\cos\phi)\nonumber\\
&&\times\delta(p_{2z}-k\sin\theta\sin\phi),\label{case3}
\end{eqnarray}
where $V$ is the volume introduced to obtain right dimension and $k$ is taken to be a constant. The first, second and third cases correspond to $P_\|/P_\perp = \infty,~0,$ and $1$, respectively.

The interaction rate, the drag coefficient and $\hat{q}$ of the charm quark for case 1 with $|\overline{M}|^2=1$ are given by

\begin{eqnarray}
\frac{dN^{\rm coll}}{dt}= \frac{\sqrt{E_1^2E_2^2-m_1^2m_2^2}}{16\pi V E_1E_2 (2E_1E_2+m_1^2+m_2^2)},~~~~~~~~~~\\
A=\frac{(E_1E_2+m_2^2)\sqrt{E_1^2E_2^2-m_1^2m_2^2}}{16\pi V E_1E_2(2E_1E_2+m_1^2+m_2^2)^2}~p_1,~~~~~~~~~~~\\
\hat{q}=\frac{k^2(E_1^2E_2^2-m_1^2m_2^2)^{3/2}}{48\pi V p_1E_2(2E_1E_2+m_1^2+m_2^2)^3}~~~~~~~~~~~~~~~~~\nonumber\\
\times\bigg\{\frac{2(2E_1E_2+m_1^2+m_2^2)}{k^2}+1+\frac{3(E_1E_2+m_1^2)^2}{E_1^2E_2^2-m_1^2m_2^2}\bigg\},\nonumber\\
\end{eqnarray}
which are derived in detail in the Appendix \ref{derivations}.

\begin{figure} [h!]
\centerline{
\includegraphics[width=8.6 cm]{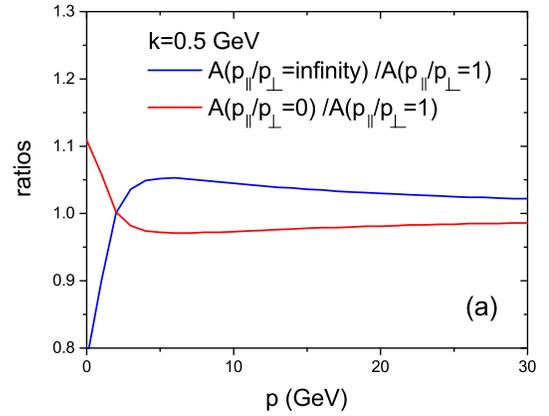}}
\centerline{
\includegraphics[width=8.6 cm]{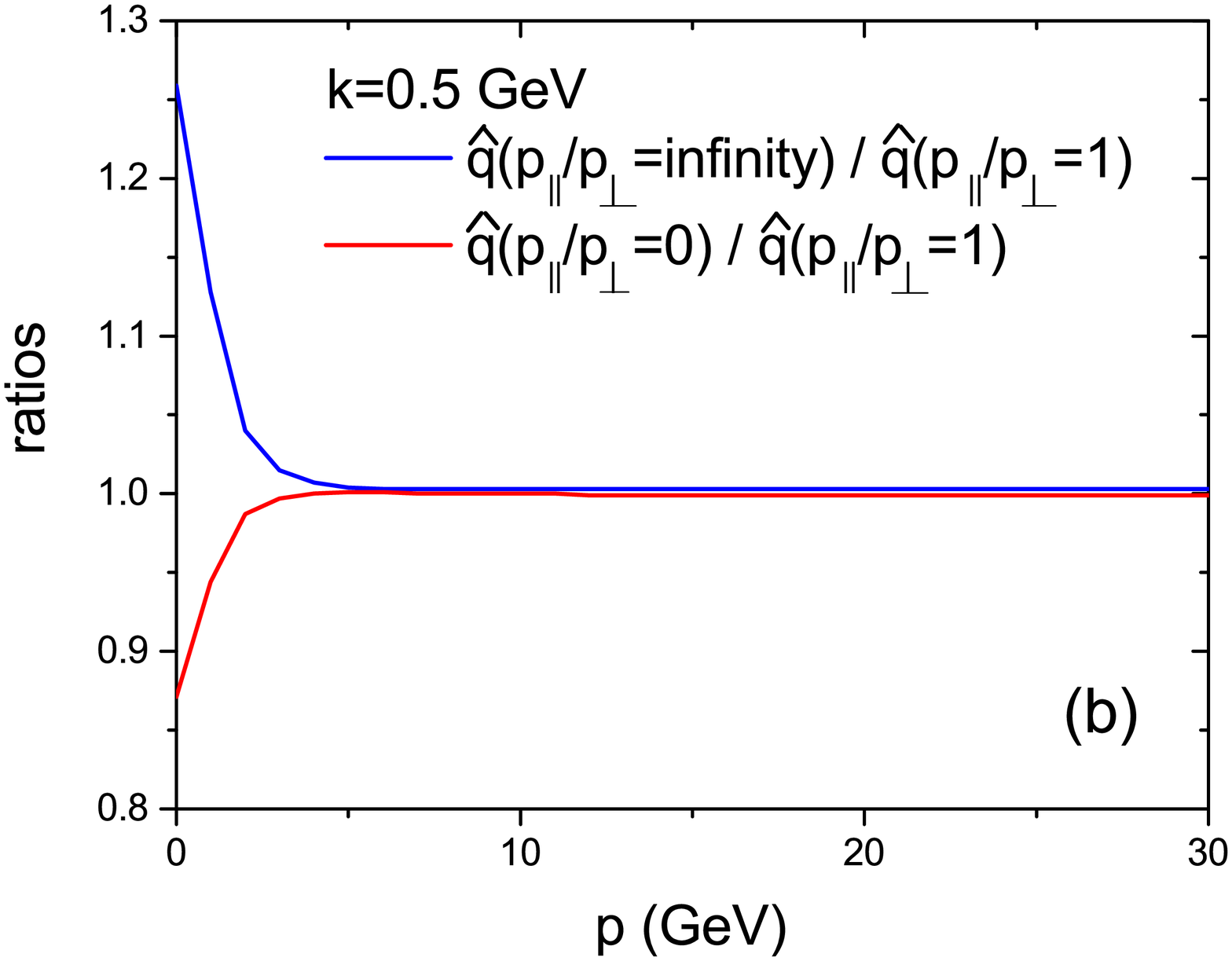}}
\caption{The ratio of the drag coefficient (upper) and $\hat{q}$ (lower) of charm quarks  at midrapidity in a QGP with anisotropic pressure  ($P_{||}/P_\bot=\infty$ and $P_{||}/P_\bot=0$) and a thermalized QGP  ($P_{||}/P_\bot=1$). We employ an isotropic scattering ($|\overline{M}|^2=1$) and $p_2$ is given by Eqs.~(\ref{case1})-(\ref{case3}) with $k$ being 0.5 GeV. The masses of the charm quark and the scattering partners are taken as 1.5 and 0.5 GeV, respectively. The initial charm quark moves in transverse direction.} \label{simple}
\end{figure}

Figure~\ref{simple} shows the ratio of the drag coefficient (top) and of $\hat{q}$ (bottom) for extremely anisotropic pressures and for isotropic pressure. The former is given by Eqs.~(\ref{case1})-(\ref{case2}) and the latter by Eqs.~(\ref{case3}), assuming isotropic scattering ($|\overline{M}|^2=1$). The masses of the charm quark and its scattering partner are taken to be 1.5 and 0.5 GeV, respectively, and the momentum of the heat bath particle is assumed to be 0.5 GeV. This gives  $k=$ 0.5 GeV in Eqs.~(\ref{case1})-(\ref{case3}).

If the charm quark has a small momentum, it will be like a static particle and the direction of momentum after scattering will follow that of scattering partner.
Therefore, in the case 1 ($P_{||}/P_\bot=\infty$) the ratio of the  drag coefficients decreases at small momenta, while that of transverse diffusion coefficient or $\hat{q}$ increases, as shown in figure~\ref{simple}. On the other hand, the ratio of the drag coefficients increases, while that of $\hat{q}$ decreases in case 2 ($P_{||}/P_\bot=0$).

When the charm quark momentum increases, the ratio of the drag coefficients shows a nontrivial behavior.
It does not converge monotonously to one but increases above one in case 1 and decreases below one in case 2,  then it slowly converges to one in both cases.
On the contrary, the ratios of  $\hat{q}$ converge monotonously to one with increasing charm quark momentum.
Since both, the ratio of the drag coefficients and that of $\hat{q}$, converge to one in the limit of large charm quark momentum, one can conclude that the angular distribution of the momentum of the scattering partner is not important for  very energetic charm quarks.

\begin{figure*}[h!]
\centerline{
\includegraphics[width=7.5 cm]{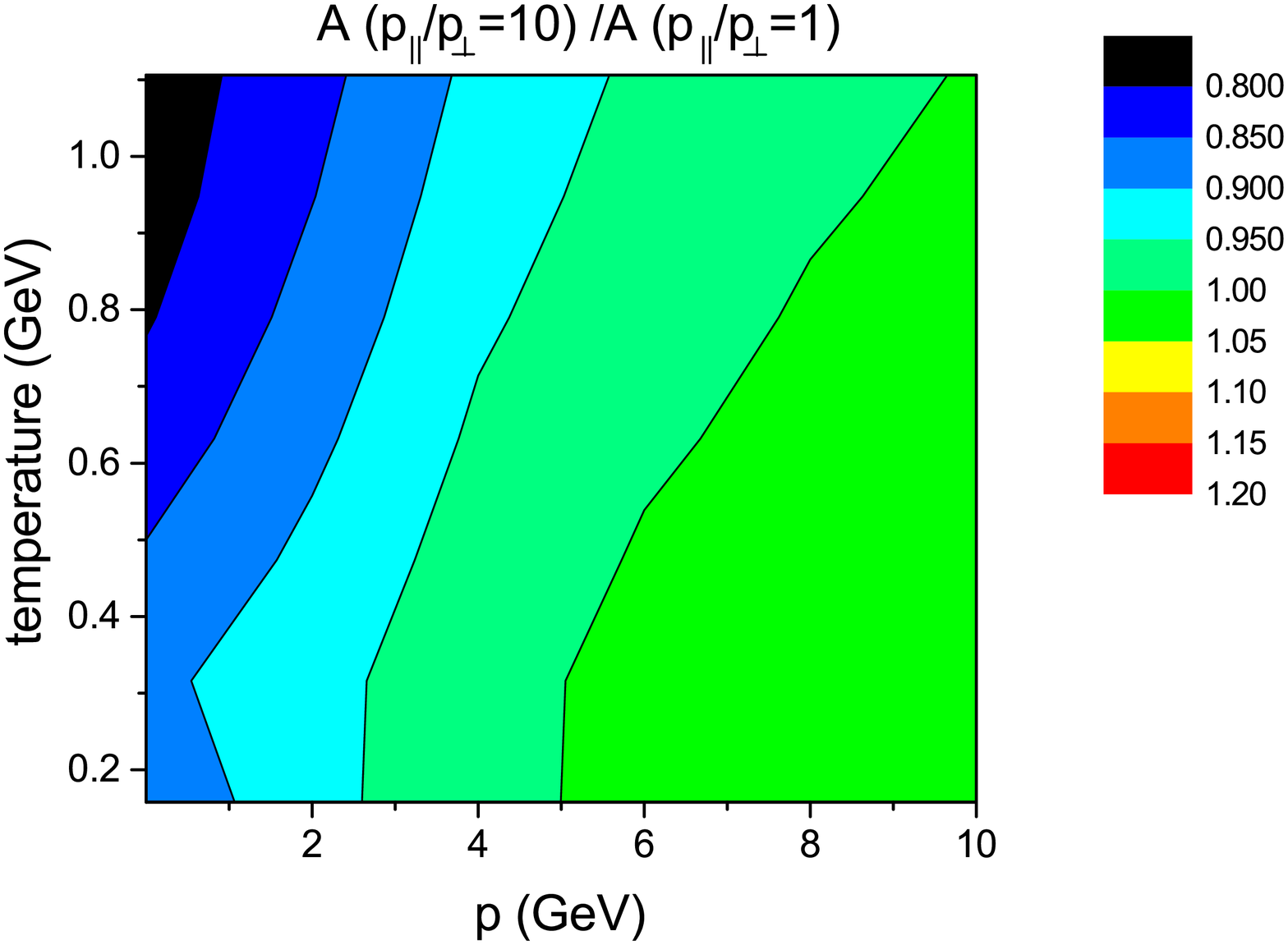}
\includegraphics[width=7.5 cm]{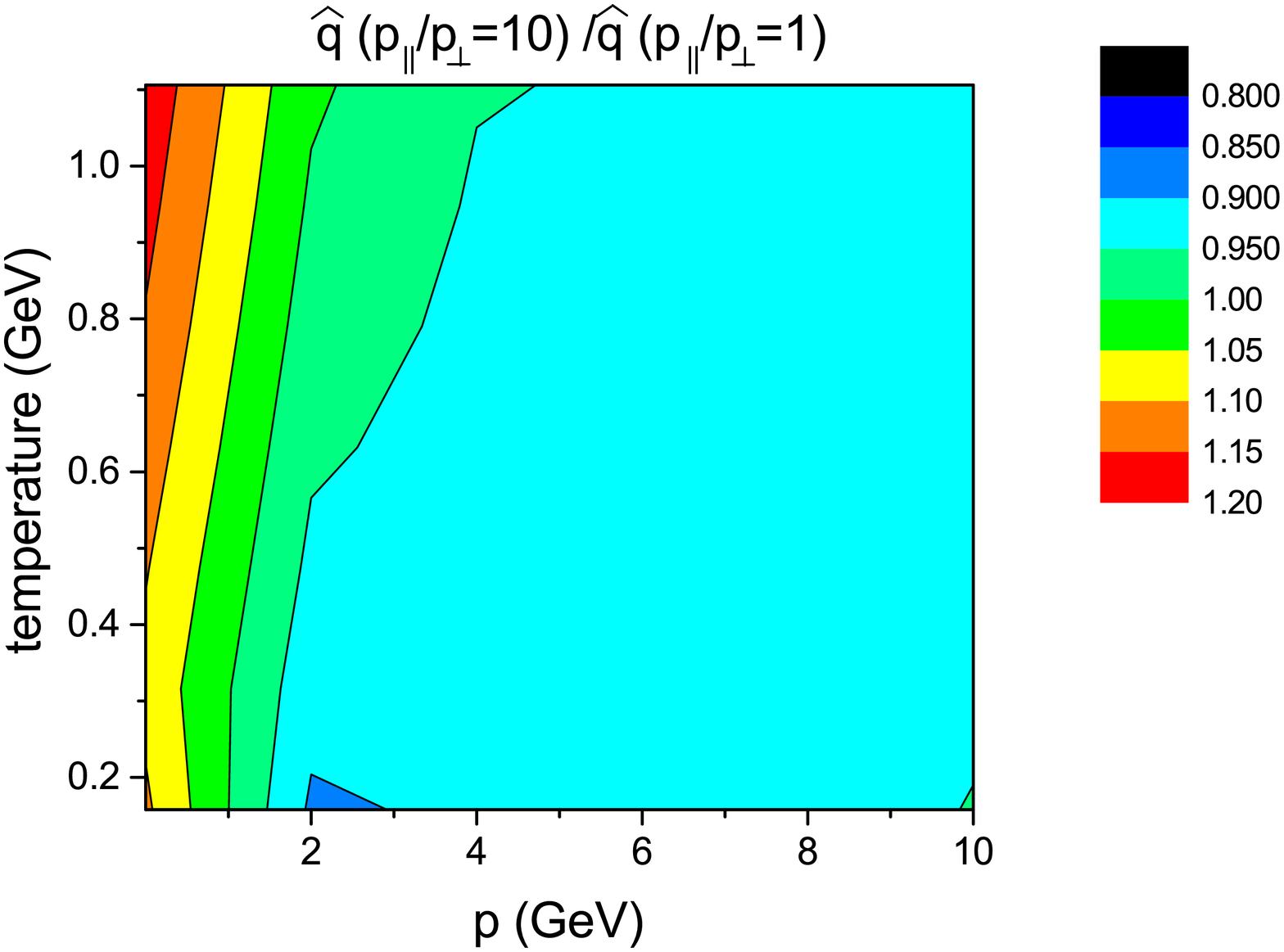}}
\centerline{
\includegraphics[width=7.5 cm]{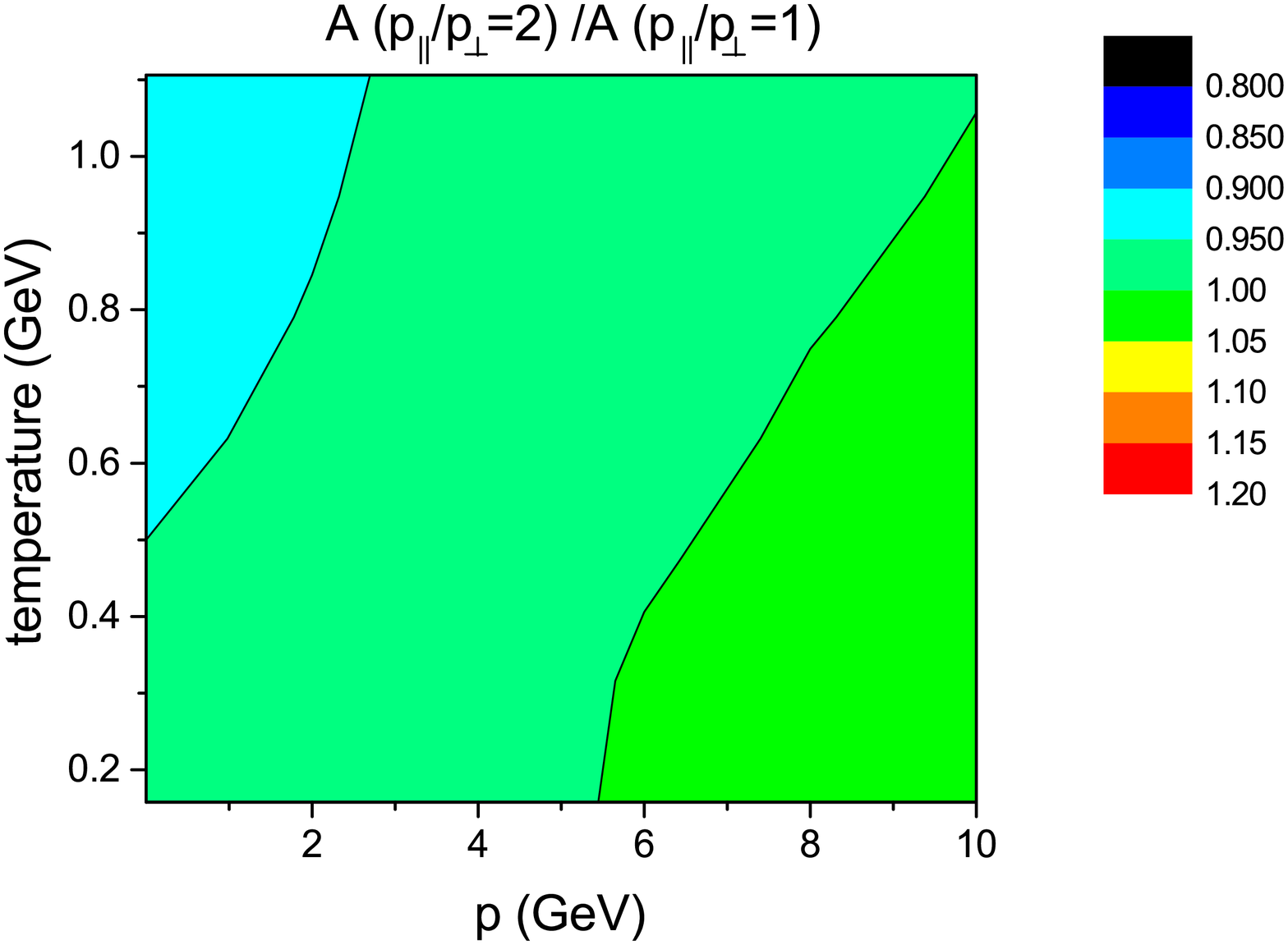}
\includegraphics[width=7.5 cm]{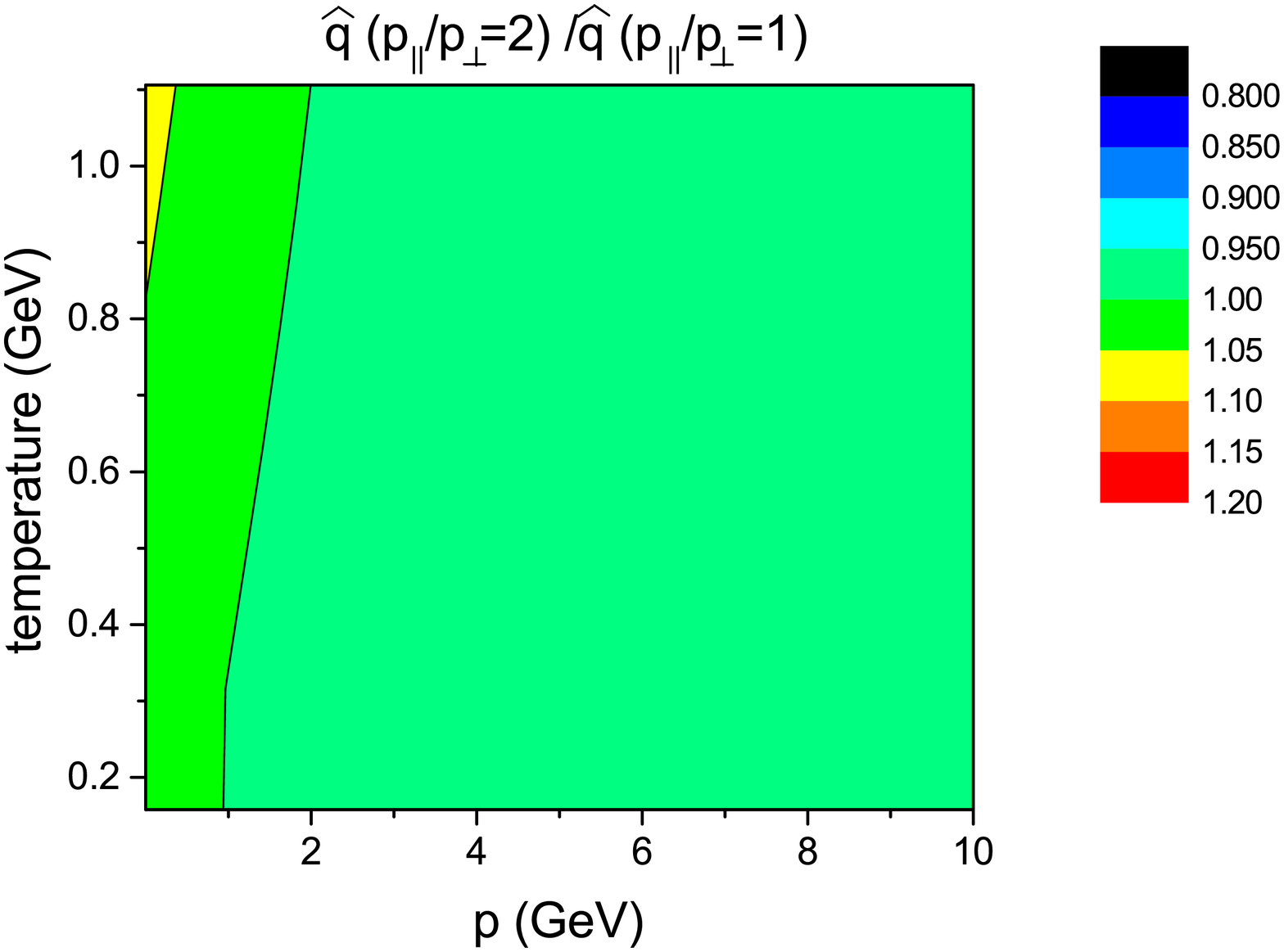}}
\centerline{
\includegraphics[width=7.5 cm]{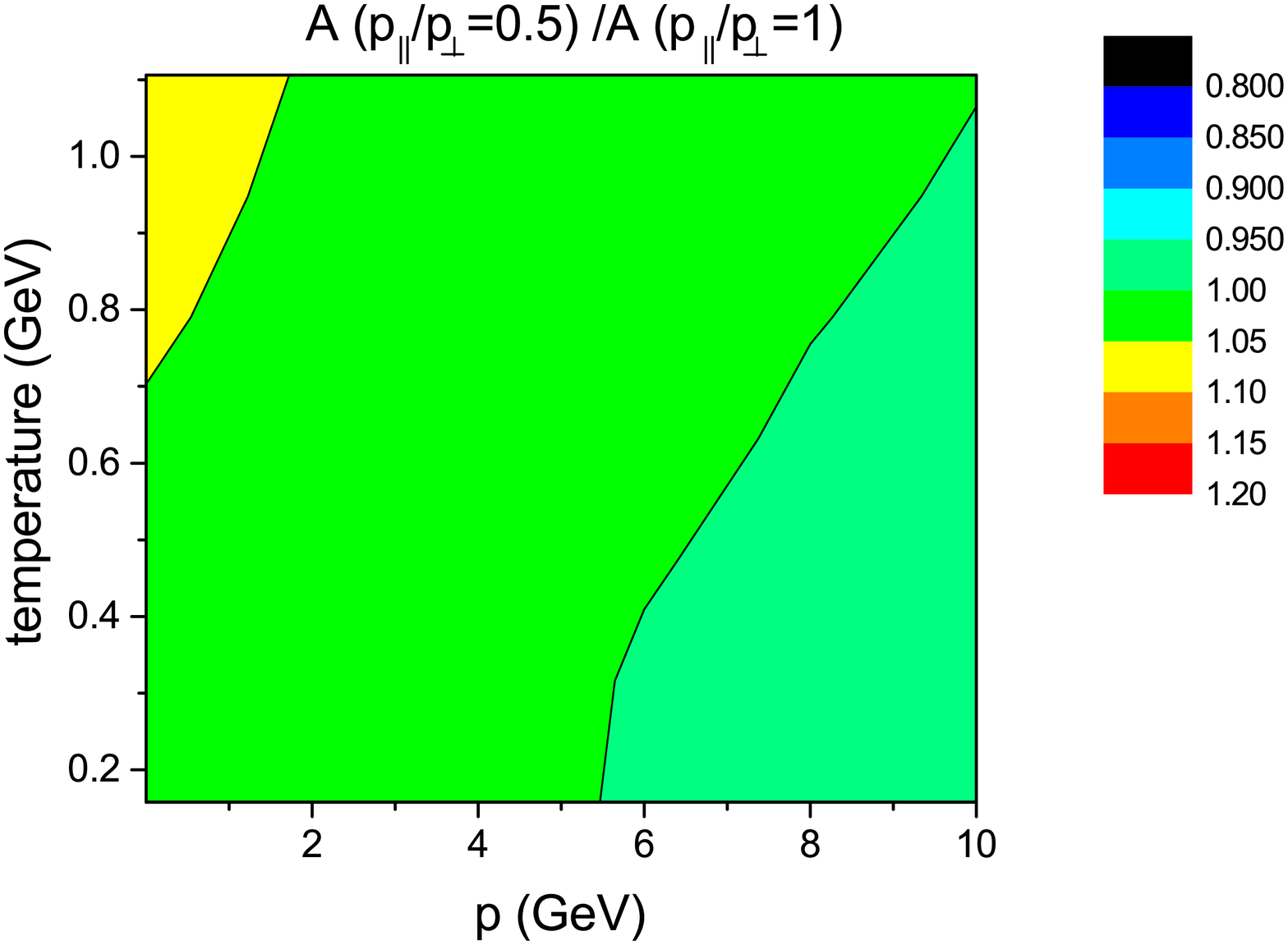}
\includegraphics[width=7.5 cm]{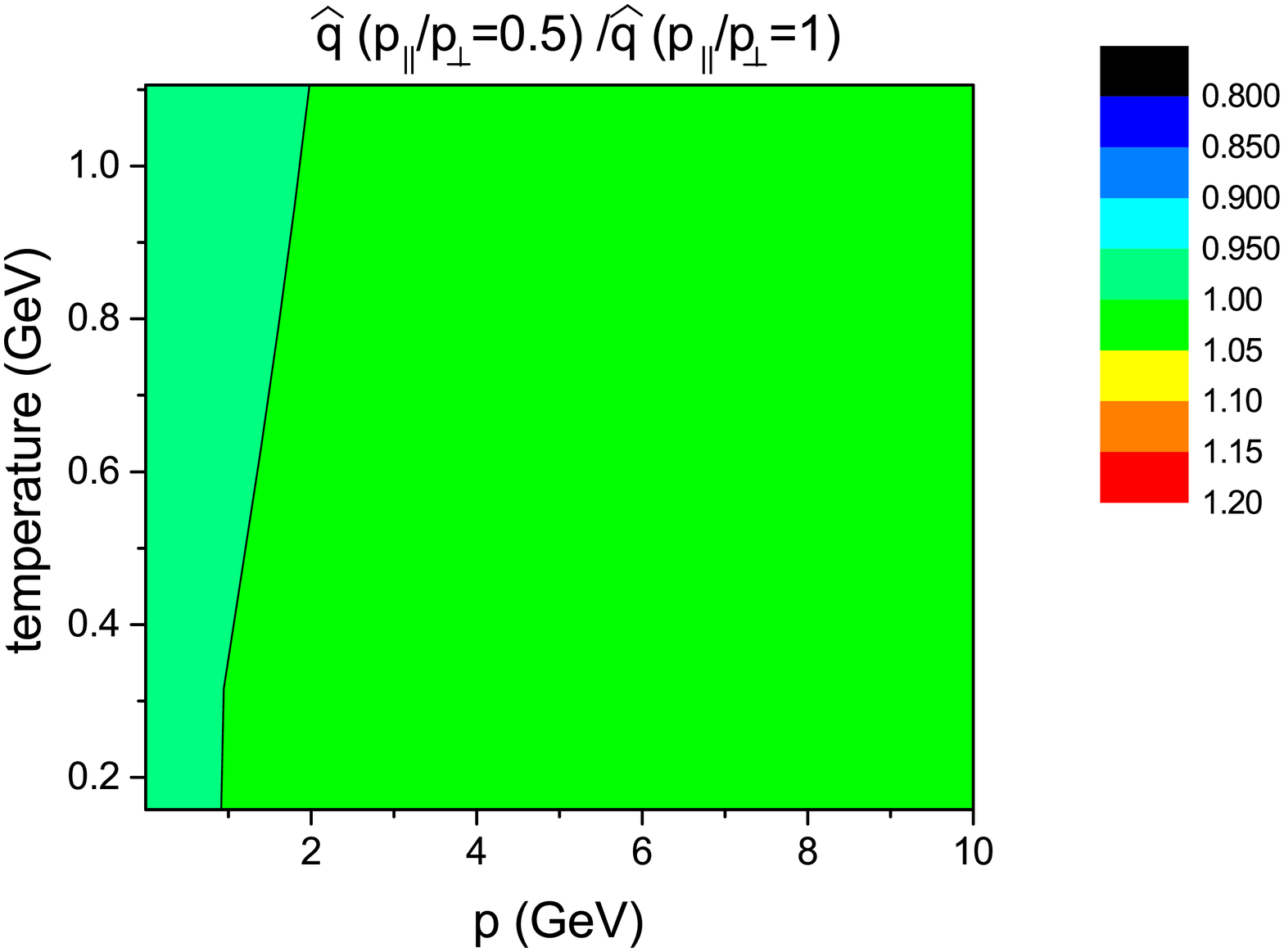}}
\centerline{
\includegraphics[width=7.5 cm]{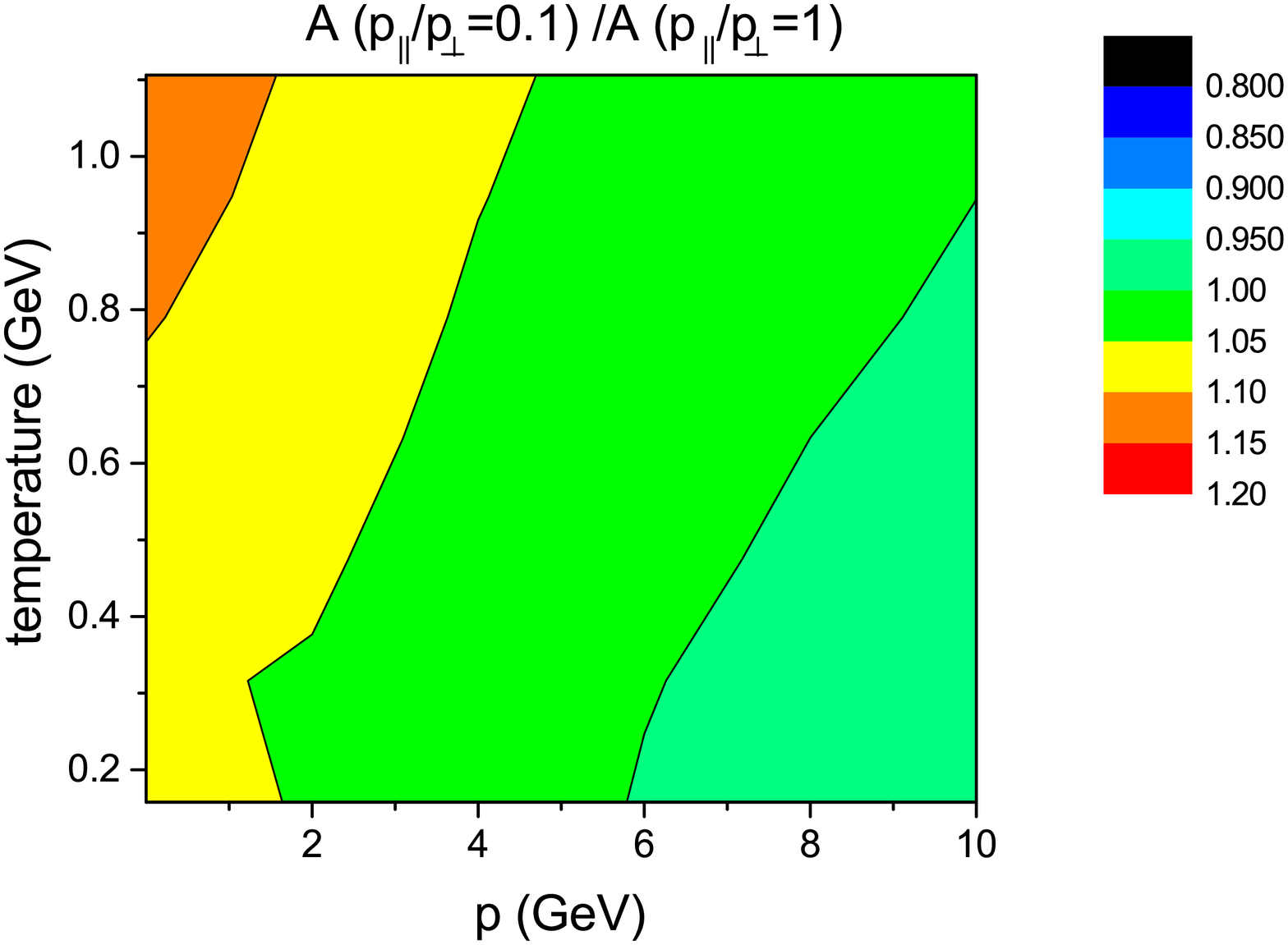}
\includegraphics[width=7.5 cm]{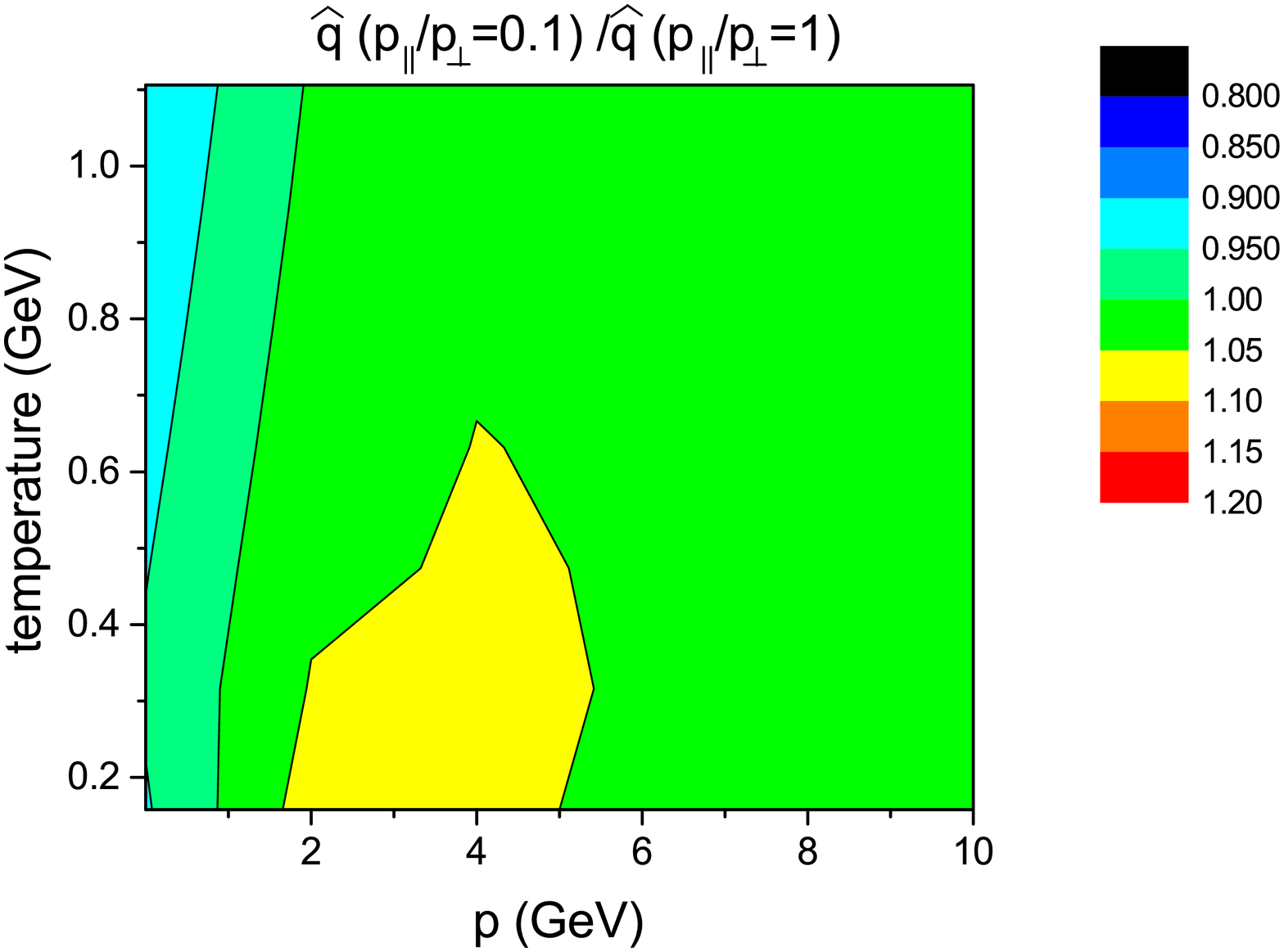}}
\caption{The ratio of drag coefficients (left) and $\hat{q}$ (right) of charm quarks at midrapidity with longitudinal pressure being 10, 2, 0.5, and 0.1 times transverse pressure and the corresponding coefficients for isotropic pressure as functions of temperature and charm quark momentum from the DQPM. The initial charm quark momentum is in transverse direction.}
\label{anisotropic}
\end{figure*}

Figure \ref{anisotropic} shows the ratios of the drag coefficients and of $\hat{q}$ of charm quarks for an anisotropic pressure and for an isotropic pressure as a function of the temperature and the charm quark momentum, employing DQPM.
One can see that the ratio of the drag coefficients has a similar behavior with increasing charm quark momentum as we observed in figure~\ref{simple} (a).
With increasing temperature the lines of a constant ratio move to larger charm quark momenta, because a larger parton momentum due to a higher temperature can be rescaled with a larger charm quark momentum.

The ratios of $\hat{q}$ are similar to those in figure~\ref{simple} (b) at low charm momentum.
However, they do not monotonously converge to one, but cross one with increasing charm momentum before converging finally to one.
This difference comes from the non-isotropic scattering. Isotropic scattering is assumed in figure~\ref{simple}, while the more realistic scattering angle distribution from PHSD is taken in figure~\ref{anisotropic}, where the scattering is more forward peaked with increasing charm quark momentum.

We conclude from figure~\ref{anisotropic} that for charm quarks with a small momentum at high temperature the modifications of the transport coefficients in a non-equilibrium QGP amounts to at most 20\% compared to the transport coefficients in an equilibrium QGP. At lower temperatures the modifications are smaller. The QGP is only for a short time  at such high temperatures. Therefore it will be difficult to see this difference in the experimental data.

\subsection{Scenario II: non-equilibrium in kinetic energy}\label{kinetic-E}

In this subsection we consider the second scenario where a non-equilibrium medium is modeled by varying the  kinetic energy. In
models which assume thermal equilibrium, for example hydrodynamics, particle spectra and multiplicities are calculated using a grand canonical  distribution.
On the other hand, transport models do not assume thermal equilibrium and particle multiplicities and momentum distributions are microscopically calculated through particle propagation and scattering in the medium. In other words, particle multiplicities and momentum distributions are dynamically generated in these simulations.

For example, the PHSD transport approach defines a local temperature through the local energy density together with the equation-of-state from lQCD~\cite{PHSD,PHSDrhic}.
The temperature defines the densities and masses in the DQPM. The same energy density can, however, be obtained by a higher average kinetic energy of the partons and a reduced parton density or by an increased parton density and a lower average kinetic energy as compared to the values obtained in complete thermal equilibrium.


To model this non-equilibrium feature, we introduce an artificial temperature $T^*$ which determines the average kinetic energy of particles by introducing the distribution function
\begin{eqnarray}
f_i(p)=\frac{D_{i}}{\exp(E/T^*)\pm 1}\times\frac{\epsilon(T)}{\epsilon(T^*)},
\label{tstar}
\end{eqnarray}
where $D_{i}$ is degeneracy factor of the particle $i$ and $\epsilon(T)$ is energy density at the temperature of $T$,
\begin{eqnarray}
\epsilon(T)=\sum_{i=q,\bar{q},g}D_{i}\int \frac{d^3p}{(2\pi)^3} \frac{E}{\exp(E/T)\pm 1}.
\end{eqnarray}
Eq.~(\ref{tstar}) describes the situation in which the momentum distribution of the partons  is characterized  by the temperature $T^*$ whereas the  energy density, $\epsilon(T)$, corresponds to the  temperature $T$:
\begin{eqnarray}
\sum_{i=q,\bar{q},g}D_{i}\int \frac{d^3p}{(2\pi)^3} E f_i(p)=\epsilon(T^*)\times\frac{\epsilon(T)}{\epsilon(T^*)}=\epsilon(T).
\end{eqnarray}

In other words, the  local energy density does not change, but the parton number density and the average kinetic energy change.
If $T^*>T$, partons have a larger kinetic energy but are less in number compared to those in thermal equilibrium.
In the other case, $T^*<T$, partons have less kinetic energy but are more abundant.

From Eq.~(\ref{tstar}) one can calculate the particle number density in the non-equilibrium situation from that in equilibrium:

\begin{eqnarray}
n^{neq}(T)=\sum_{i=q,\bar{q},g}D_{i}\int \frac{d^3p}{(2\pi)^3}f_i(p)=n^{eq}(T^*)\frac{\epsilon(T)}{\epsilon(T^*)},
\end{eqnarray}
and obtains the ratio
\begin{eqnarray}
\frac{n^{neq}(T)}{n^{eq}(T)}=\frac{n^{eq}(T^*)}{n^{eq}(T)}\times\frac{\epsilon(T)}{\epsilon(T^*)}.
\label{ratio-numden}
\end{eqnarray}

We note that the ratio of the number density in the DQPM is more complicated than Eq.~(\ref{ratio-numden}) because of the spectral function.

\begin{figure} [h!]
\centerline{
\includegraphics[width=8.6 cm]{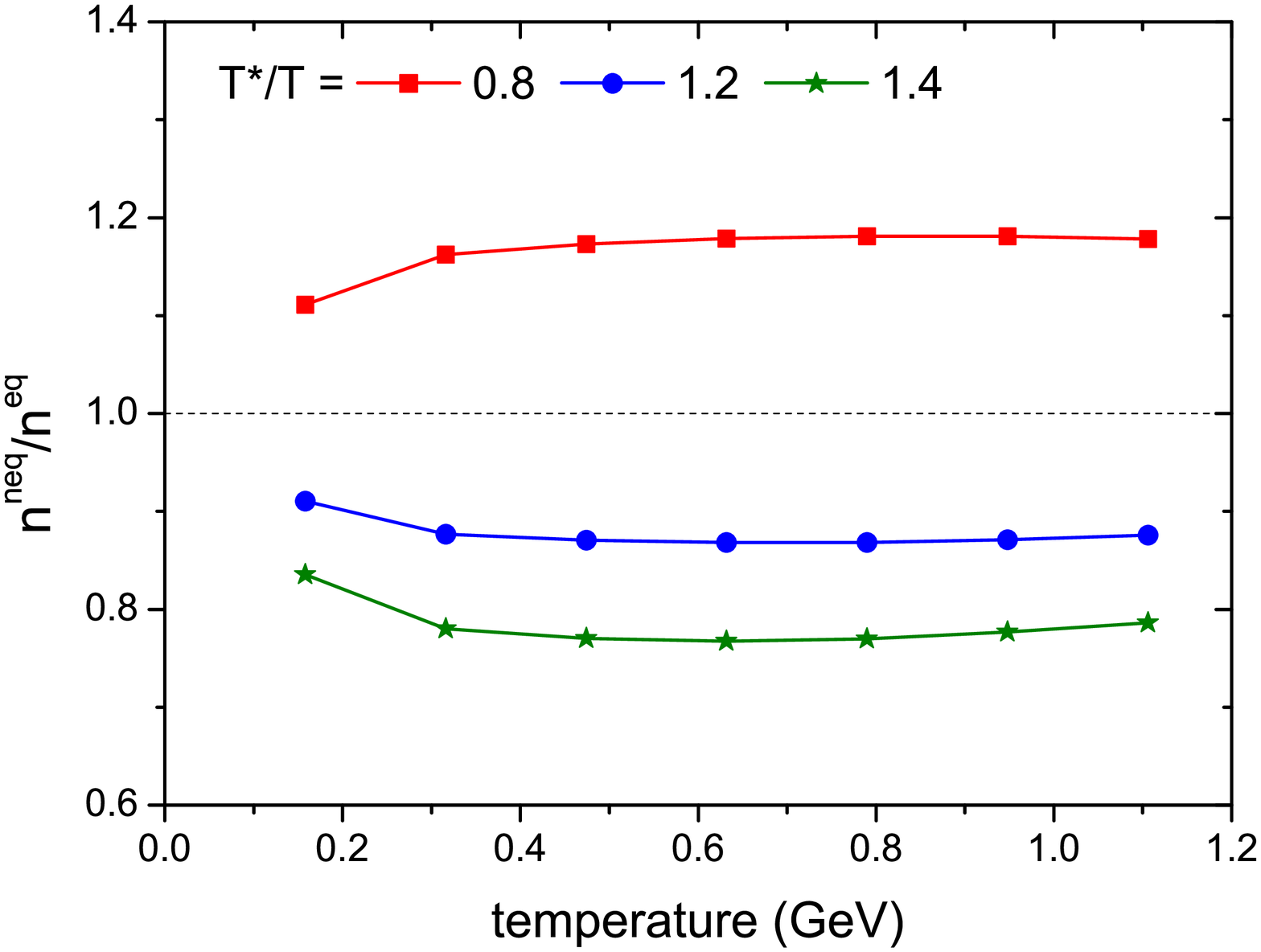}}
\caption{The ratios of parton number densities  in  non-equilibrated  and equilibrated matter for $T^*/T=$ 0.8, 1.2, and 1.4 from the DQPM, assuming that the energy density is that of equilibrated matter. } \label{num-ratio}
\end{figure}

The ratio of the particle number densities in non-equilibrium and that in equilibrium from the DQPM as a function of temperature and for a couple of temperature ratios, $T^*/T$, is shown in figure~\ref{num-ratio}.
As mentioned, for $T^*/T<1$ the particle number density increases and the particles have smaller kinetic energies than in thermal equilibrium.
On the other hand, the particle number density is smaller for $T^*/T>1$ and the kinetic energy increases to keep the energy density unchanged.
The ratio is closer to one near $T_c$, because in the DQPM the parton mass increases as the temperature approaches $T_c$.

\begin{figure} [h!]
\centerline{
\includegraphics[width=8.6 cm]{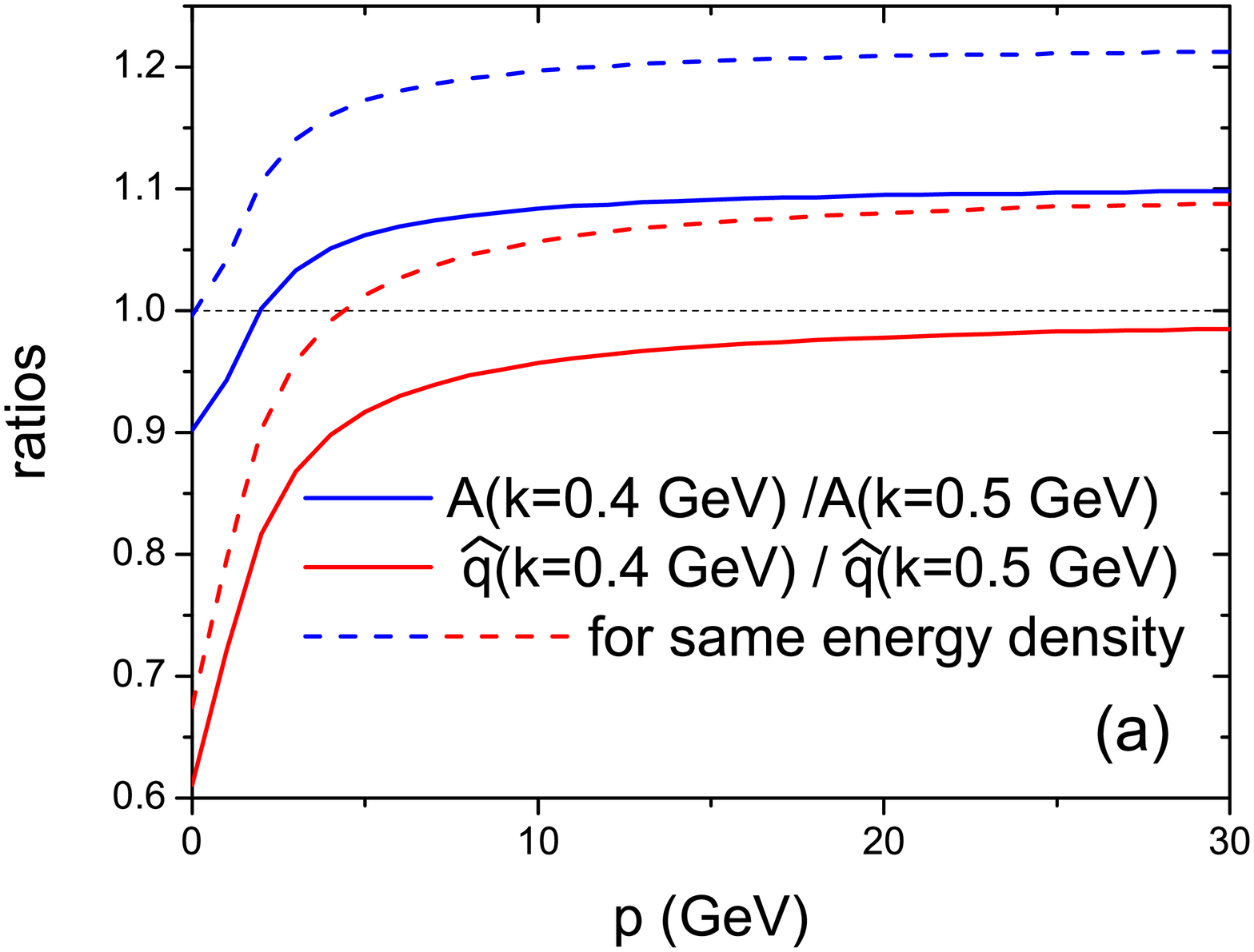}}
\centerline{
\includegraphics[width=8.6 cm]{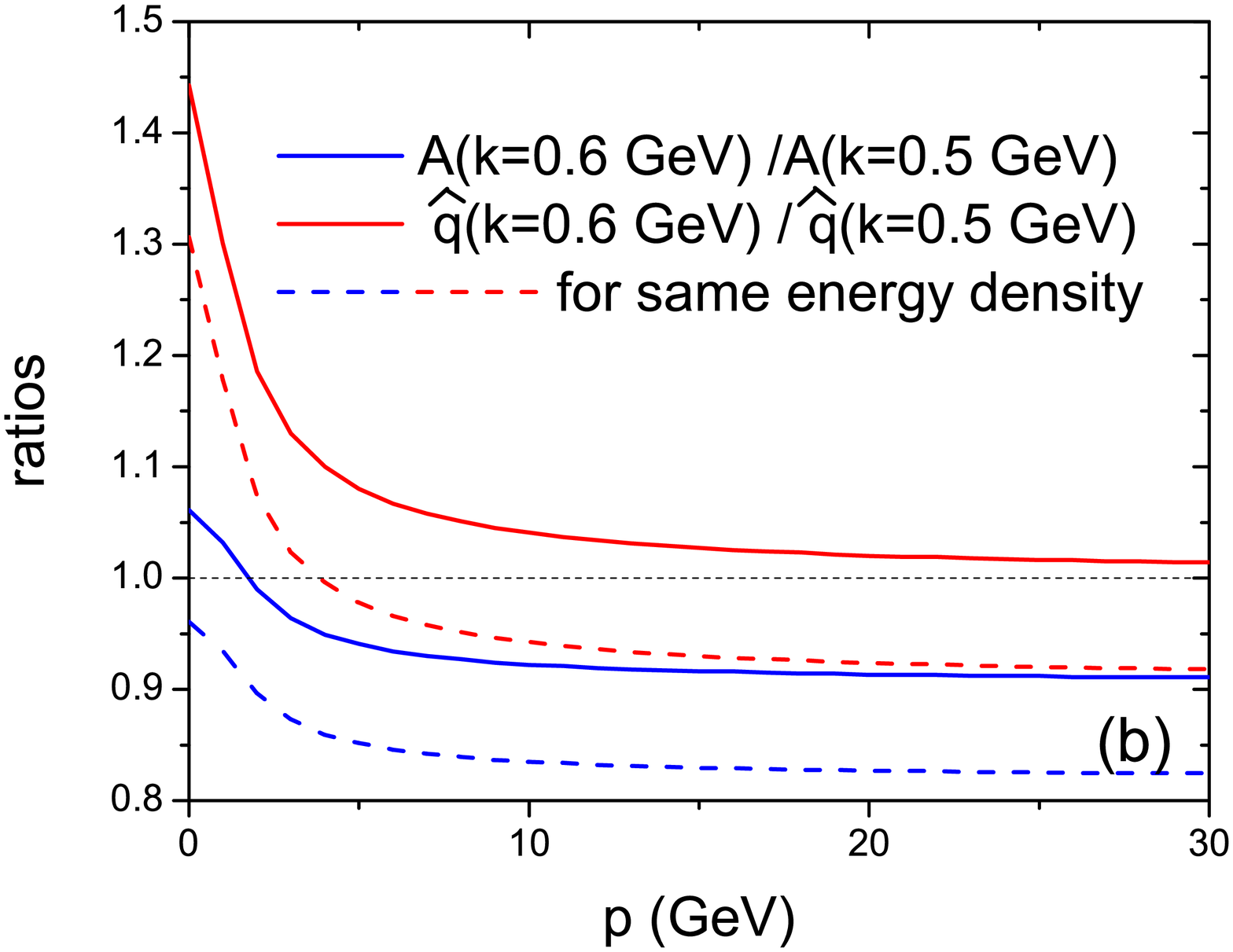}}
\caption{The drag coefficient and $\hat{q}$ of charm quarks which scatters off a parton whose momentum is $k=0.4$ GeV (upper panel) or $k=0.6$ GeV (lower panel) in Eq.~(\ref{case3}) divided by the same transport coefficients for $k=0.5$ GeV as a function of the charm quark momentum. The parton mass is taken as 0.5 GeV and isotropic scattering ($|\overline{M}|^2=1$) is assumed.
Dashed lines are the ratios multiplied by energy ratios to make the same energy density of the matter.}
\label{temperature}
\end{figure}

Before calculating the drag coefficient and $\hat{q}$ of charm quarks in the DQPM, one can roughly estimate both in the simple case of Eq.~(\ref{case3}), where the parton momentum is isotropic and has the constant norm $k$ which is taken as $0.4$, $0.5$, and $0.6$ GeV. We assume a parton mass of 0.5 GeV.
The ratios of the drag coefficient and $\hat{q}$ for $k=0.4$ GeV and $0.6$ GeV and that for $k=0.5$ GeV are shown in figure~\ref{temperature}.

One might expect that transport coefficients of heavy quarks are large if the scattering partner has a large kinetic energy.
This is true for the drag coefficient and $\hat{q}$ at small momentum of the charm quark as shown in figure~\ref{temperature}.
As the charm quark momentum increases, the ratio of the drag coefficients becomes larger than one for $k=$ 0.4 GeV, and  smaller than one for $k=$ 0.6 GeV.
While the ratio of $\hat{q}$ converges to one with increasing charm quark momentum, the ratio of drag coefficient does not.

In the previous subsection we showed that the drag coefficient of charm quarks does not depend on the angular distribution of the parton momentum in the limit of a large charm quark momentum. Therefore, Eq.~(\ref{case3}) can be reduced to Eq.~(\ref{case1}), where the parton momentum $\vec{k}$ is either $k_z$ or $-k_z$.
From Eq.~(\ref{drag3}) of Appendix~\ref{derivations} we see that the asymptotic form of the drag coefficient is given by
\begin{eqnarray}
\lim_{p_1\rightarrow\infty} A=\frac{1}{64\pi V E_2},
\label{limit3}
\end{eqnarray}
where $p_1$ is the charm quark momentum, from which the ratio of the drag coefficients for different charm quark momenta is given by
\begin{eqnarray}
\lim_{p_1\rightarrow\infty} \frac{A(p_2^\prime)}{A(p_2)}=\frac{E_2}{E_2^\prime},
\label{asymptotic}
\end{eqnarray}
where $(E_2,p_2)$ and $(E_2^\prime,p_2^\prime)$ are two different energy-momentum vectors of the scattering partner of the charm quark. For example, we display in figure~\ref{temperature} (a)
this ratio for $p_2=0.4$ GeV and $p_2^\prime=0.5$ GeV.
The inverse proportionality of the drag coefficient to the energy of the scattering partner originates from the scattering cross section
\begin{eqnarray}
\lim_{E_1\rightarrow \infty}\sigma=\frac{1}{32\pi E_1E_2},
\label{cs-inftyb}
\end{eqnarray}
which is derived in Appendix~\ref{derivations}. As the energy of the scattering partner, $E_2$, increases, the scattering cross section decreases.
One can see that the asymptotic ratio of Eq.~(\ref{asymptotic})
\begin{eqnarray}
\lim_{p_1\rightarrow\infty} \frac{A(p_2=0.4~ {\rm GeV})}{A(p_2=0.5~ {\rm GeV})}&\simeq& 1.1,\nonumber\\
\lim_{p_1\rightarrow\infty} \frac{A(p_2=0.6~ {\rm GeV})}{A(p_2=0.5~ {\rm GeV})}&\simeq& 0.91.\nonumber
\end{eqnarray}
 explains figure~\ref{temperature} very well.

On the other hand, the ratio of $\hat{q}$ is always smaller than one in figure~\ref{temperature} (a) and larger than one in figure~\ref{temperature} (b).
Both ratios converge to one in the limit of large momentum of the charm quark, because the transverse momentum squared per collision is proportional to $E_1E_2$ in this limit, as shown in Eq.~(\ref{dpt2b}) of Appendix~\ref{derivations}.
If the energy of the scattering partner, $E_2$, increases, the charm quark gains a larger transverse momentum.
However, this enhancement is exactly canceled by the reduction of scattering cross section in Eq.~(\ref{cs-inftyb}).
Therefore the asymptotic value of $\hat{q}$ does not depend on the kinetic energy of the scattering partner as derived in Appendix~\ref{derivations}:

\begin{eqnarray}
\lim_{E_1\rightarrow \infty}\hat{q}=\frac{1}{96\pi V}.
\end{eqnarray}

The dashed lines in figure~\ref{temperature} are ratios of transport coefficients multiplied by energy density ratios to have the same energy density of the matter, for example, $\sqrt{m^2+k^2}/\sqrt{m^2+k^{\prime 2}}=\sqrt{0.5^2+0.5^2}/\sqrt{0.5^2+0.4^2}$ is multiplied to the upper panel.

\begin{figure*}[h!]
\centerline{
\includegraphics[width=7.5 cm]{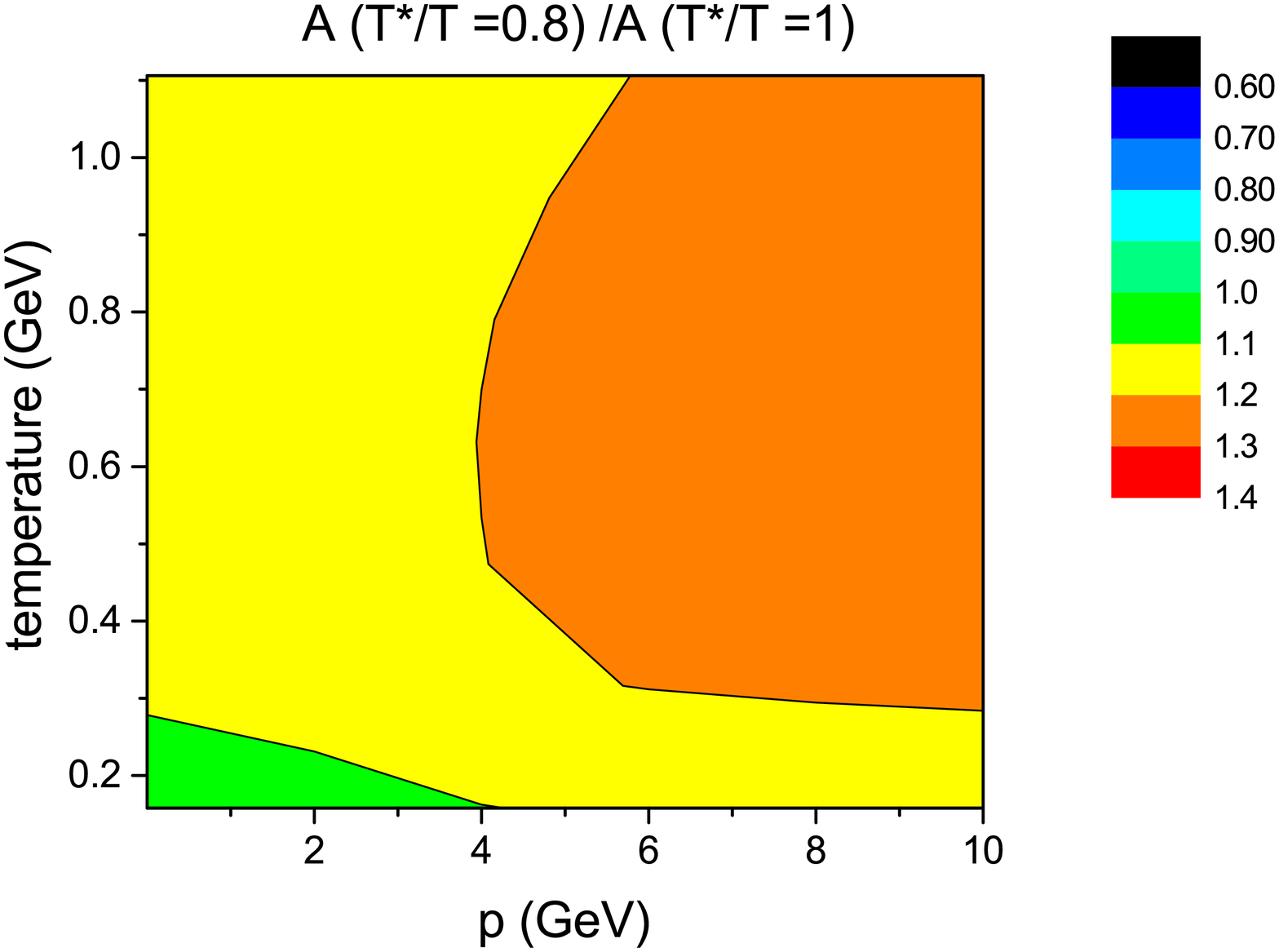}
\includegraphics[width=7.5 cm]{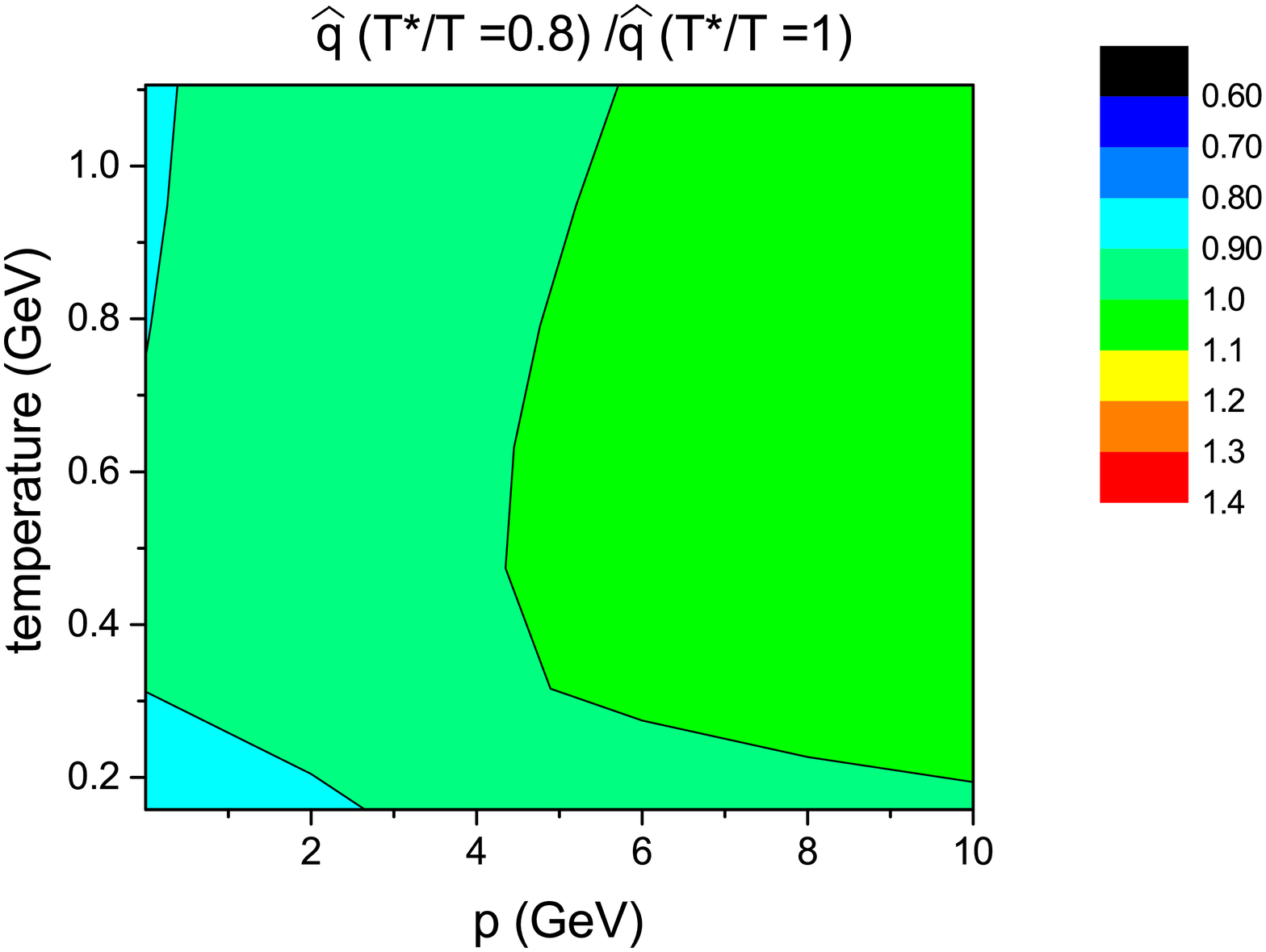}}
\centerline{
\includegraphics[width=7.5 cm]{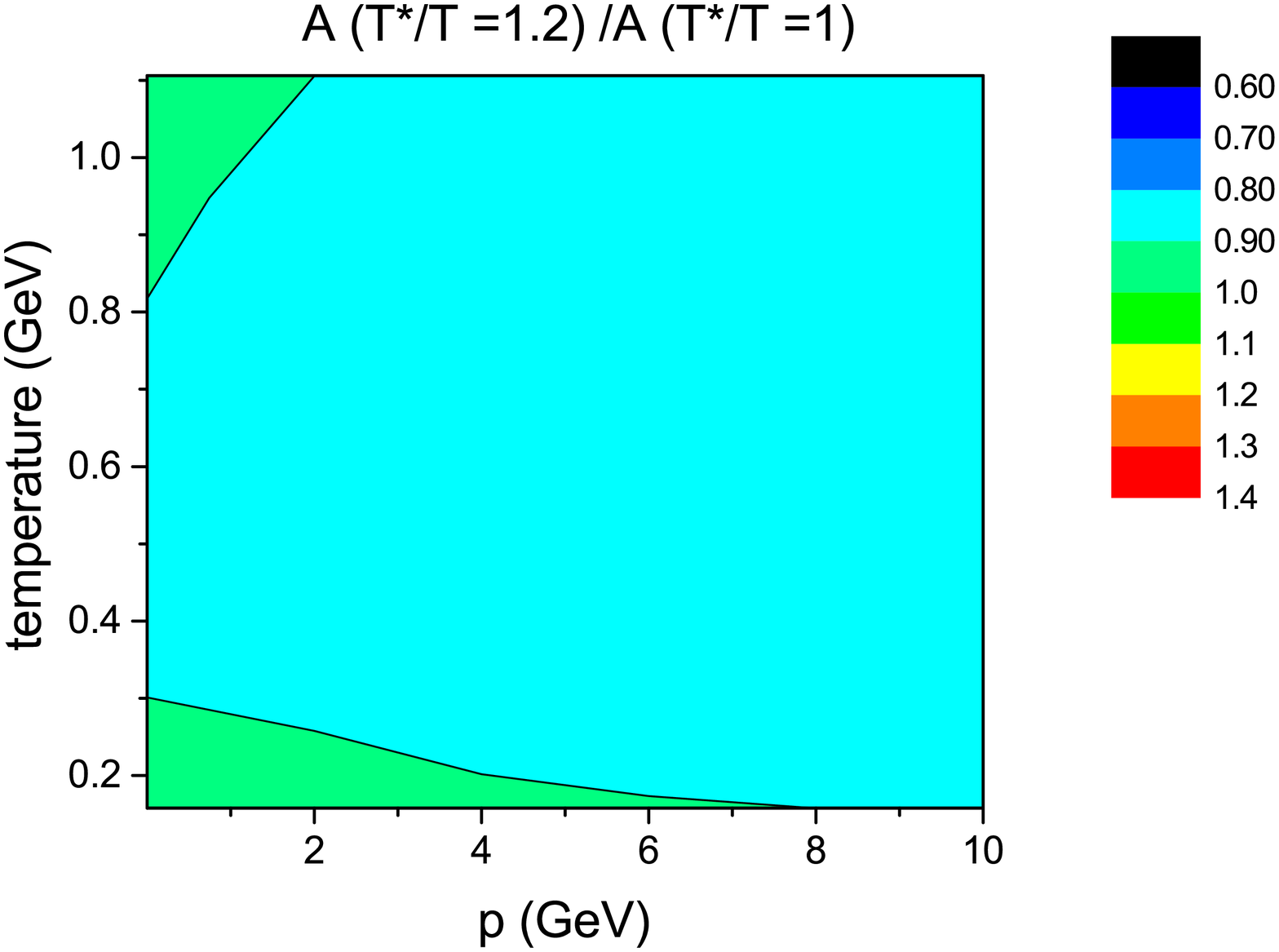}
\includegraphics[width=7.5 cm]{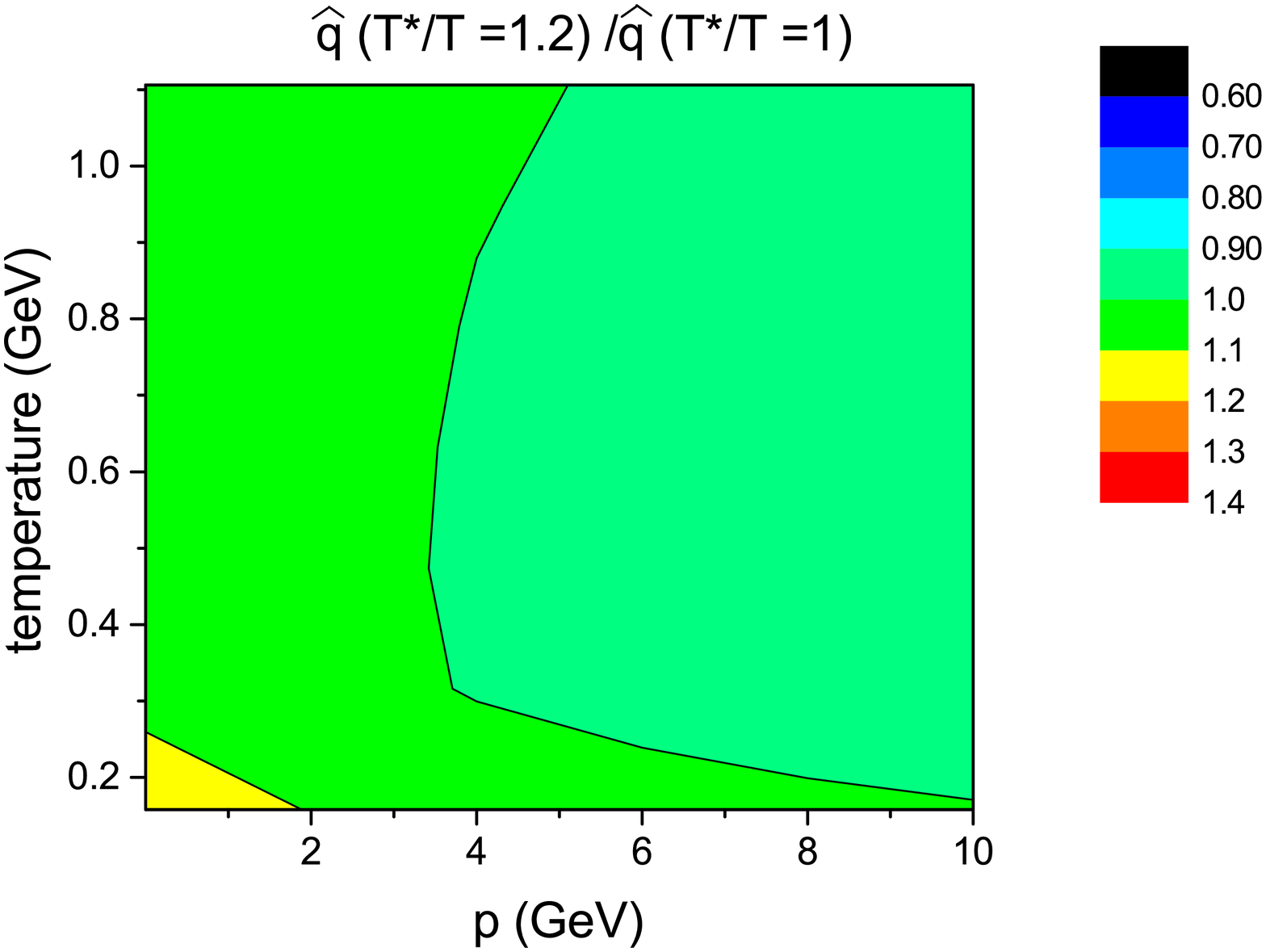}}
\centerline{
\includegraphics[width=7.5 cm]{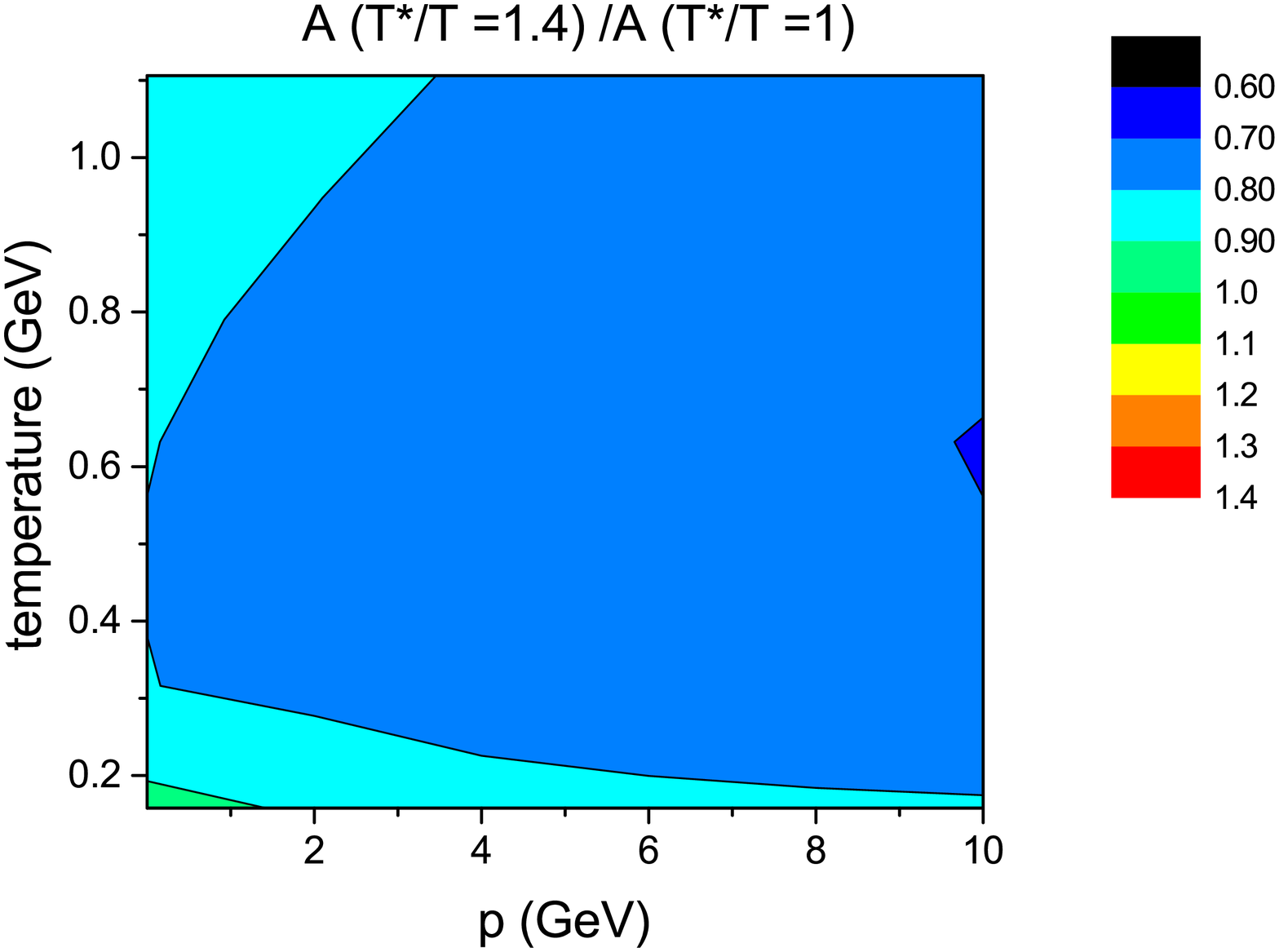}
\includegraphics[width=7.5 cm]{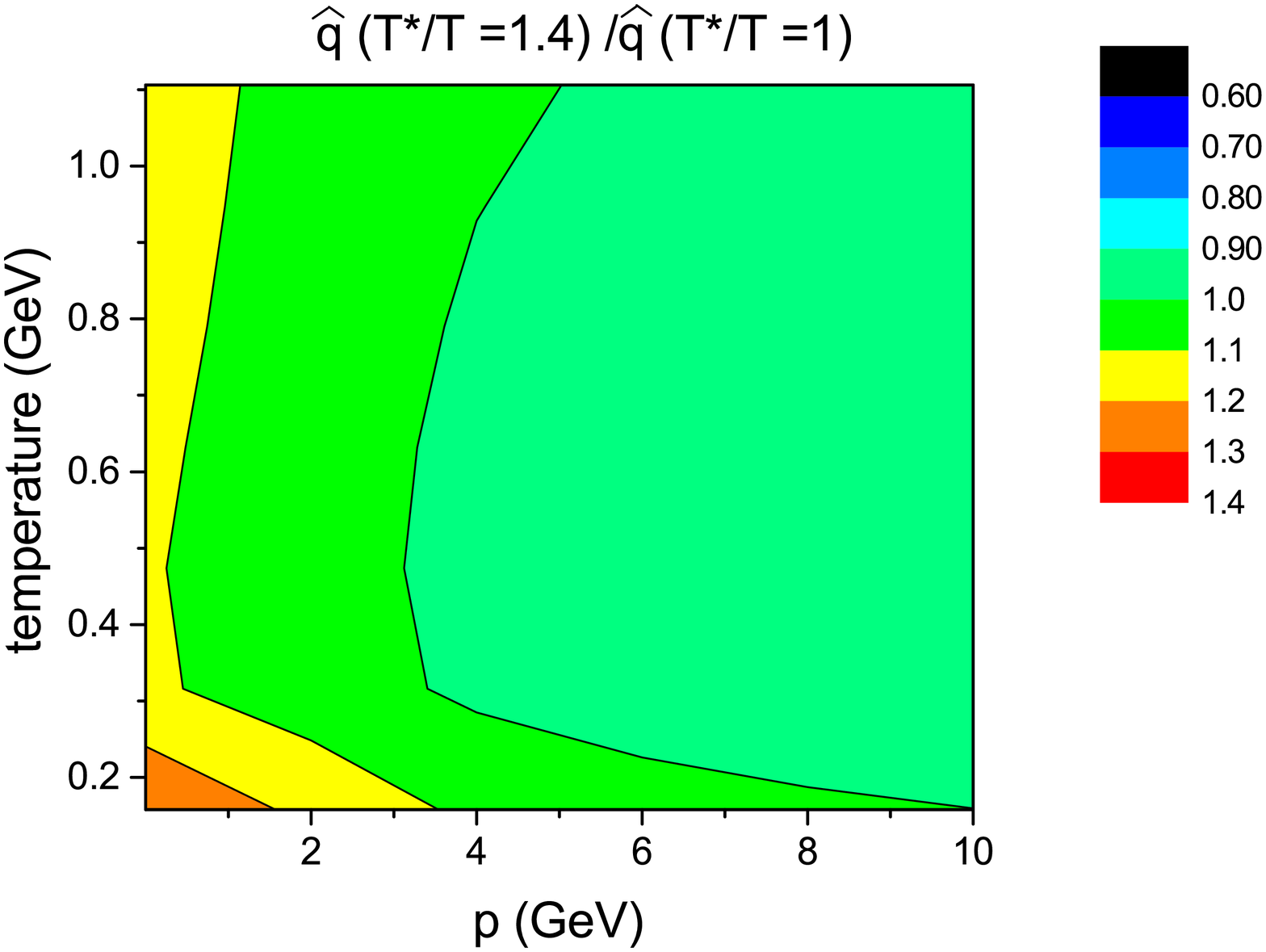}}
\caption{The ratio of the drag coefficient (left) and $\hat{q}$ (right) of a charm quark with the kinetic energy being not in equilibrium and those in equilibrium for $T^*/T=$ 0.8, 1.2, and 1.4 as functions of the temperature and the charm quark momentum from the DQPM calculations.}
\label{ratiot}
\end{figure*}

To get a more realistic result, the ratios are calculated in the DQPM and shown in figure~\ref{ratiot}.
It is different from the solid lines but similar to dashed line in figure~\ref{temperature} because the ratio of particle number densities of figure~\ref{num-ratio} is involved to keep the energy density unchanged.
The transport coefficient is sensitive to the number density of partons, because it directly enters the scattering rate of heavy quarks inside the QGP.
This is one of the physical reasons why some transport coefficients do not converge to the equilibrium limit even if the heavy quark momentum is large.
This additional constrain raises the ratios of transport coefficient for $T^*/T=$ 0.8 and lowers the ratios for $T^*/T=$ 1.2 and 1.4, which is shown in figure~\ref{ratiot}.
As a consequence, the asymptotic values of both transport coefficients, the drag coefficient and $\hat{q}$, are larger than one for $T^*/T=$ 0.8 and smaller than one for $T^*/T=$ 1.2 and 1.4 in the limit of a large momentum of charm quark.
This means that for a constant energy density heavy quarks  lose more energy in matter when partons have a small kinetic energy but are numerous, and lose less energy when they have a large kinetic energy but a smaller number density.

\subsection{Scenario III: non-equilibrium in parton mass}\label{off-shell}

In the DQPM the QGP  is composed of massive quarks and gluons whose spectral function has temperature-dependent pole mass and width fitted to the equation-of-state from lattice QCD.
When the matter, produced in the heavy-ion collisions, expands, the temperature decreases and the pole mass and spectral width of the partons, which depend on temperature, are modified. The change of the parton spectral function can be realized by quasi-elastic scattering which reassigns the parton mass.
If the average time between parton scatterings is extremely short,  we employ the spectral function of the DQPM.
However, if the average time is not short, the mass distribution of partons will be different from that in thermal equilibrium.
We call this approach "non-equilibrium in parton mass" and study its influence on the transport coefficients of heavy quarks by shifting the pole mass.
As in the previous subsection (scenario II), the energy density remains unchanged by  rescaling the parton number density.

\begin{figure} [h!]
\centerline{
\includegraphics[width=8.6 cm]{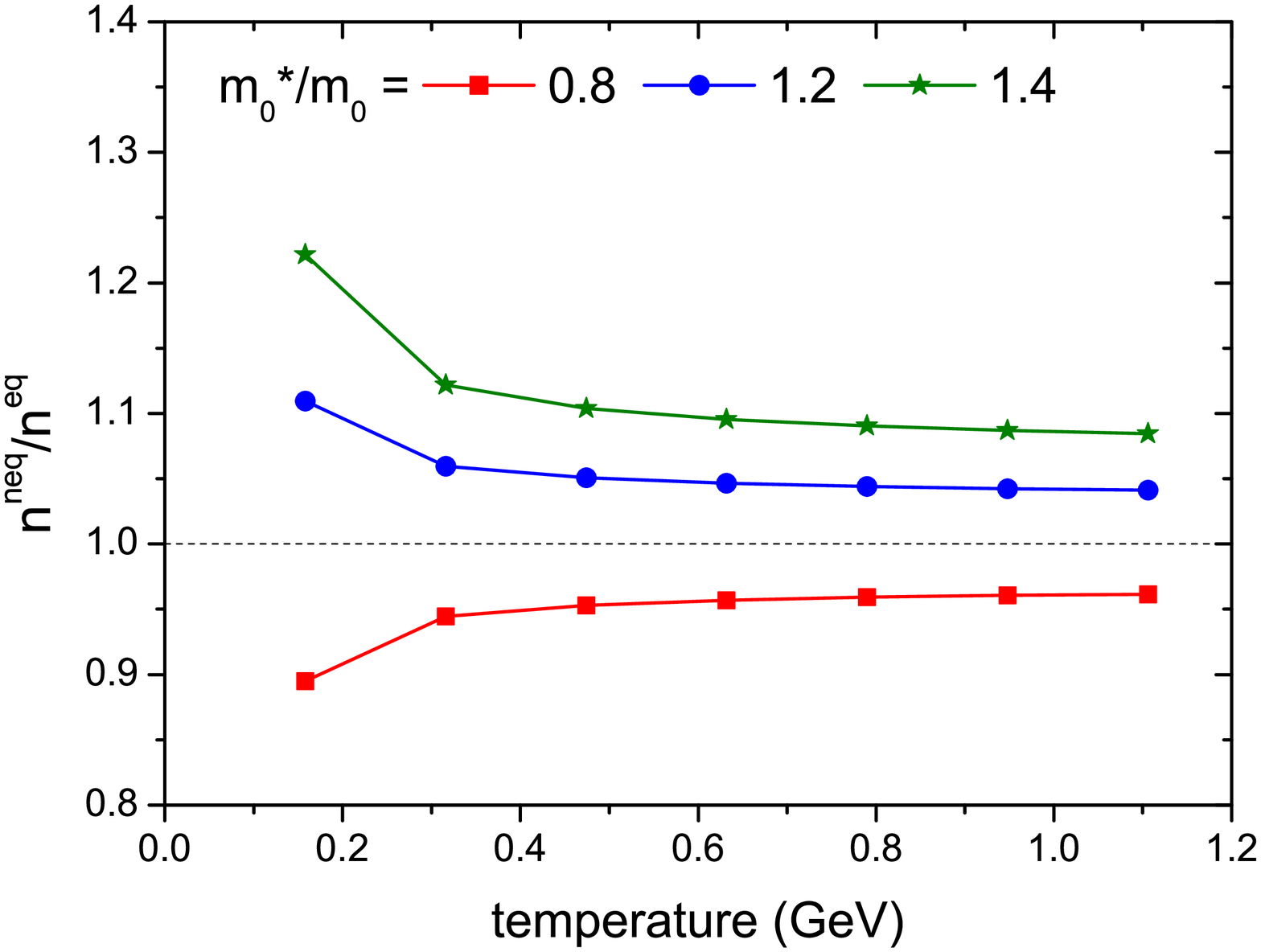}}
\caption{The ratios of the parton number density with the pole mass being 0.8, 1.2, and 1.4 times the pole mass from the DQPM and the number density without shifting the pole mass as a function of the temperature, assuming that the equilibrium and non-equilibrium energy density is the same.} \label{num-ratiom}
\end{figure}

Figure~\ref{num-ratiom} shows the ratio of the parton number density with a shifted pole mass and that without such a shift. The shifted pole masses are 0.8, 1.2, and 1.4 times the pole mass from the DQPM. The energy density without and with shift is identical.
The ratio of the number density is larger than one for a smaller a pole mass, and smaller than one for a larger pole mass.
The deviation from one increases with decreasing temperature  because of the dependence of the parton spectral function on the temperature in the DQPM.

\begin{figure} [h]
\centerline{
\includegraphics[width=8.6 cm]{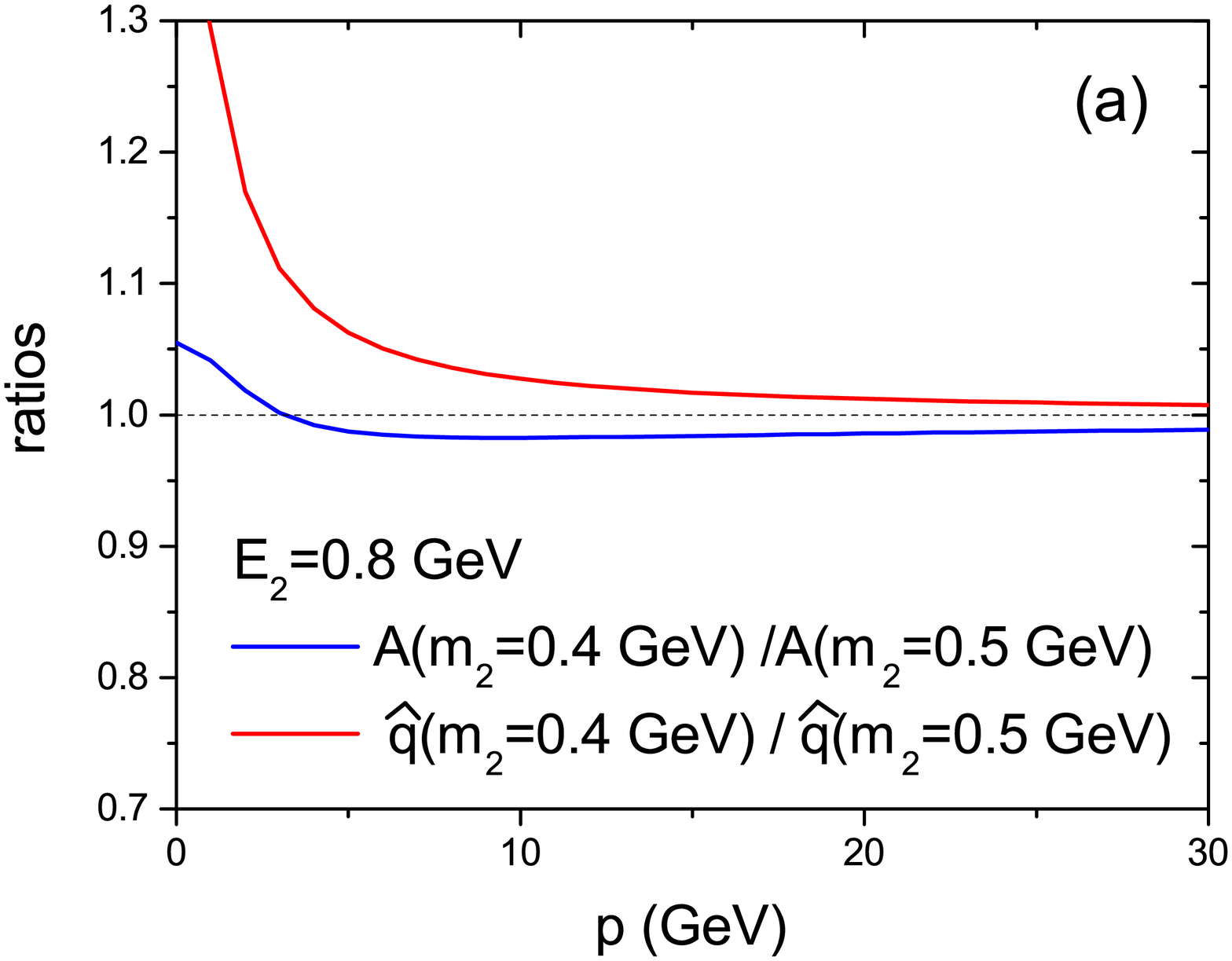}}
\centerline{
\includegraphics[width=8.6 cm]{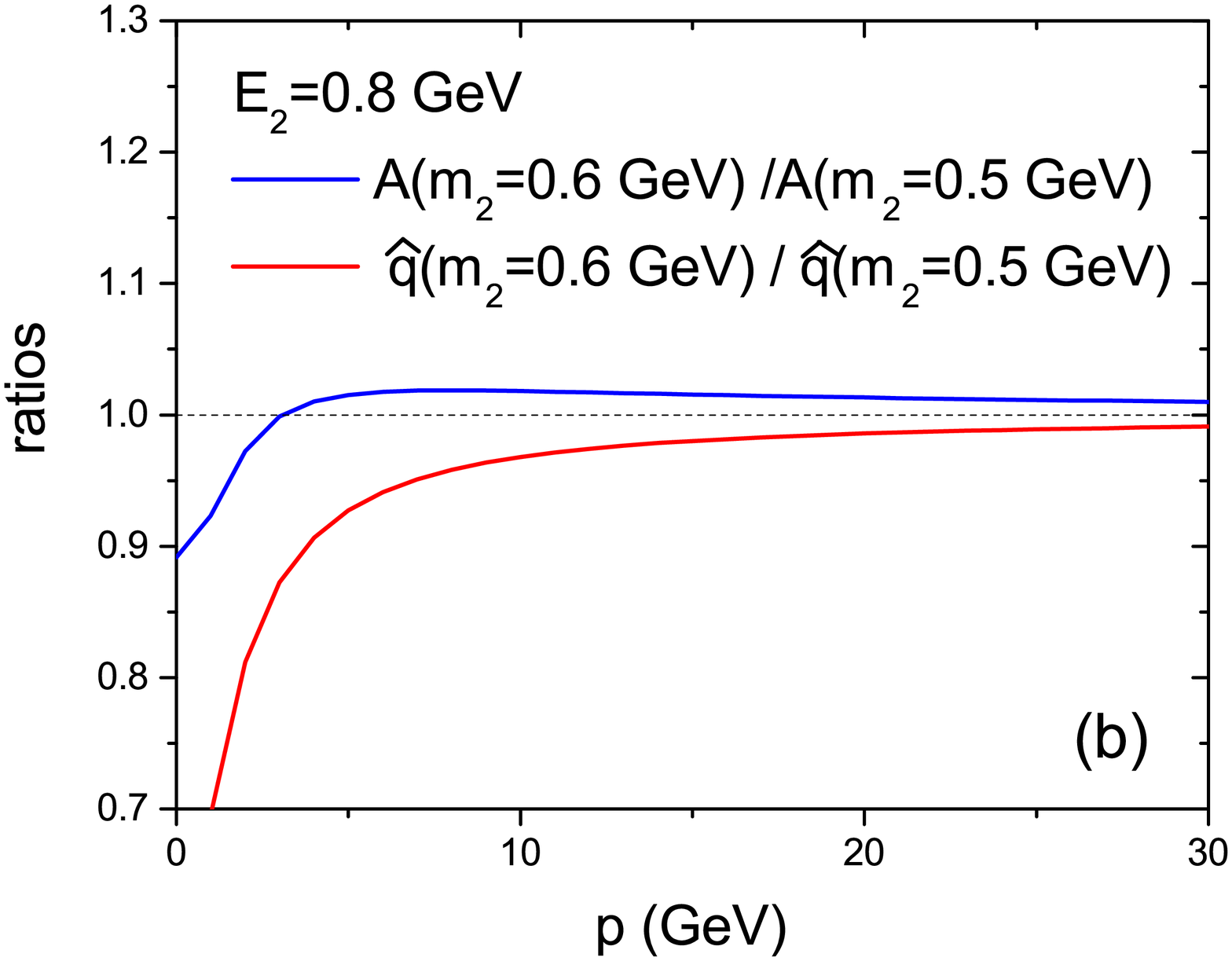}}
\caption{The ratio of the drag coefficient and $\hat{q}$ of a charm quark which scatters off a particle whose mass is 0.4 GeV (upper) and 0.6 GeV (lower) and the same coefficient for a particle mass of 0.5 GeV as a function of the charm quark momentum. The momentum distribution of the scattering partner is given by Eq.~(\ref{case3}) with $E_2=\sqrt{m_2^2+k^2}=$ 0.8 GeV and isotropic scattering ($|\overline{M}|^2=1$) is assumed.}
\label{mass}
\end{figure}

We first discuss the simple case of a momentum distribution given by $f(p_2)$ of Eq.~(\ref{case3}) where the parton momentum is isotropic with a constant magnitude $k$.
Figure~\ref{mass} shows the ratio of the drag coefficient and $\hat{q}$ of a charm quark which scatters off a parton whose mass is 0.4 GeV and 0.6 GeV and the same transport coefficient for a  mass  of 0.5 GeV  as a function of the charm quark momentum, assuming isotropic scattering ($|\overline{M}|^2=1$).
In order to come from the momentum distribution function $f(p_2)$ of Eq.~(\ref{case3}) to a thermal distribution function, we have to weight $f(p_2)$ with the Boltzmann factor $e^{-E/T}$.
Therefore, one can mimic the thermal distribution by using a common parton energy of 0.8 GeV and using the corresponding value of $k$ in figure~\ref{mass}.
One can see that at small momenta of the charm quarks both, $A$ and $\hat{q}$, for a parton mass of 0.4 GeV (0.6 GeV) are larger (smaller) than those for a parton mass of 0.5 GeV.
If the charm quark momentum increases, the ratio of the transport coefficients converges to one, regardless of the parton mass, though the ratio of drag coefficients crosses one before converging.
The convergence to one is consistent with Eqs.~(\ref{drag-infty}) and (\ref{qhat-infty}) in Appendix~\ref{derivations}, which depend on $E_2$, not on $m_2$.

\begin{figure*}[h!]
\centerline{
\includegraphics[width=7.5 cm]{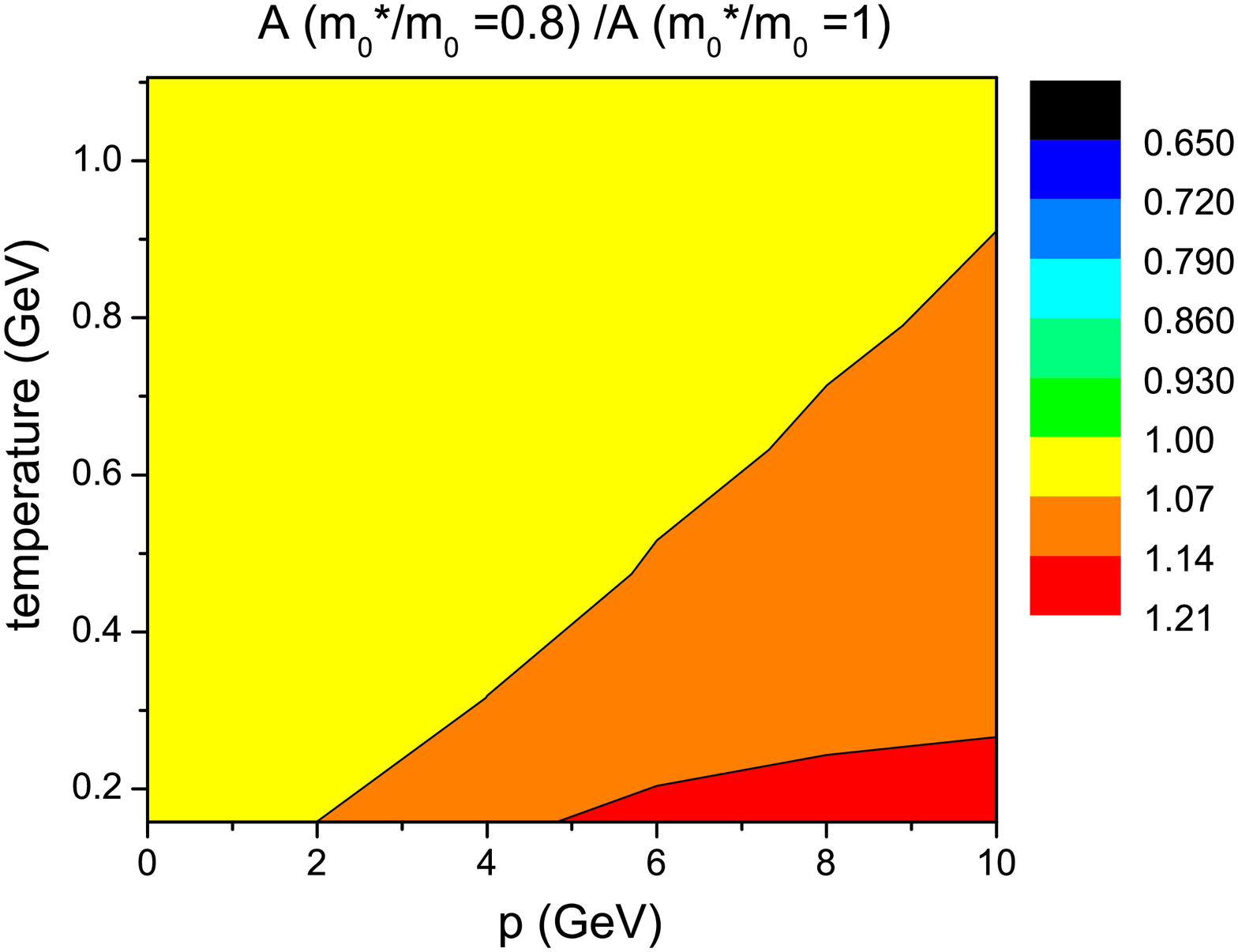}
\includegraphics[width=7.5 cm]{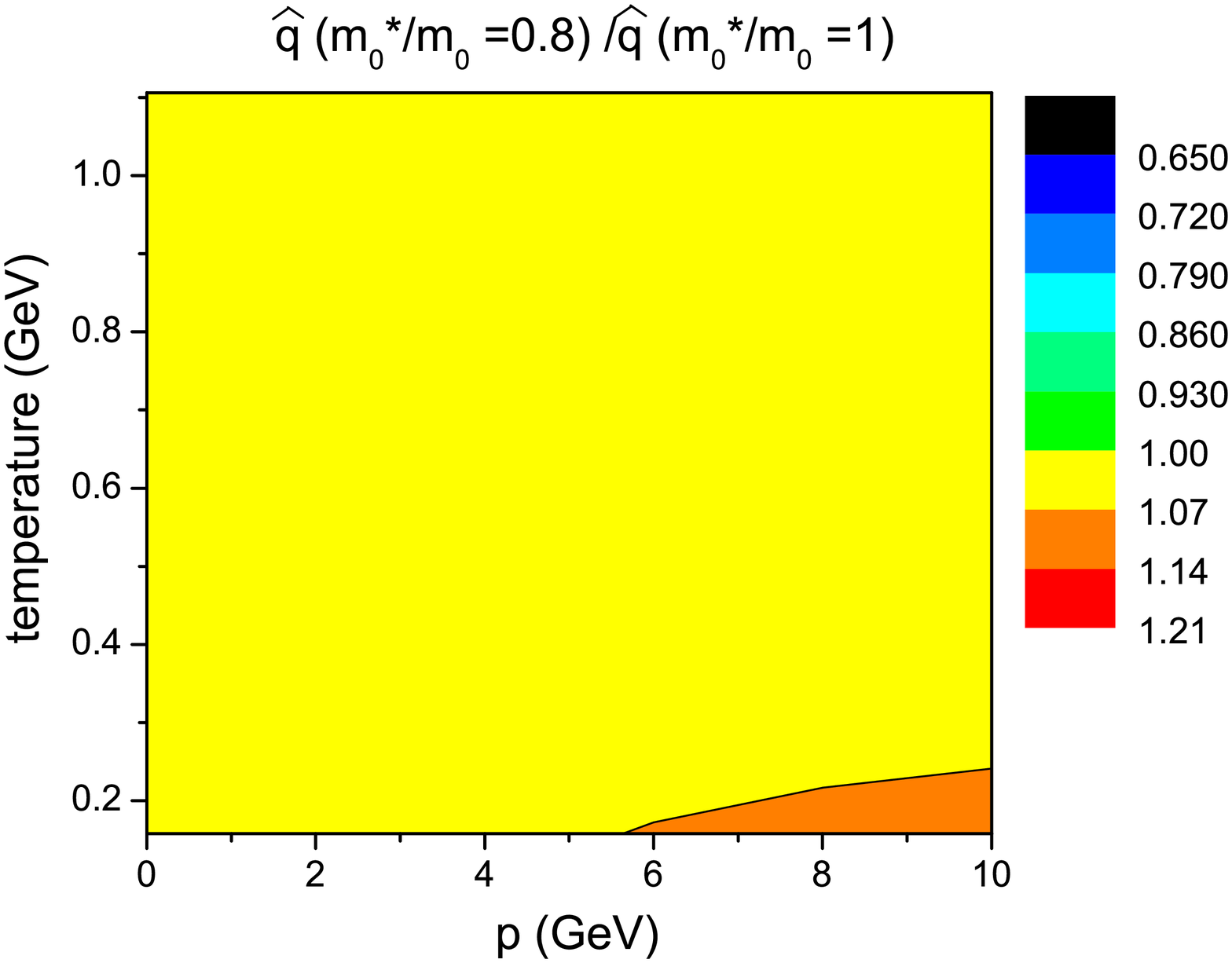}}
\centerline{
\includegraphics[width=7.5 cm]{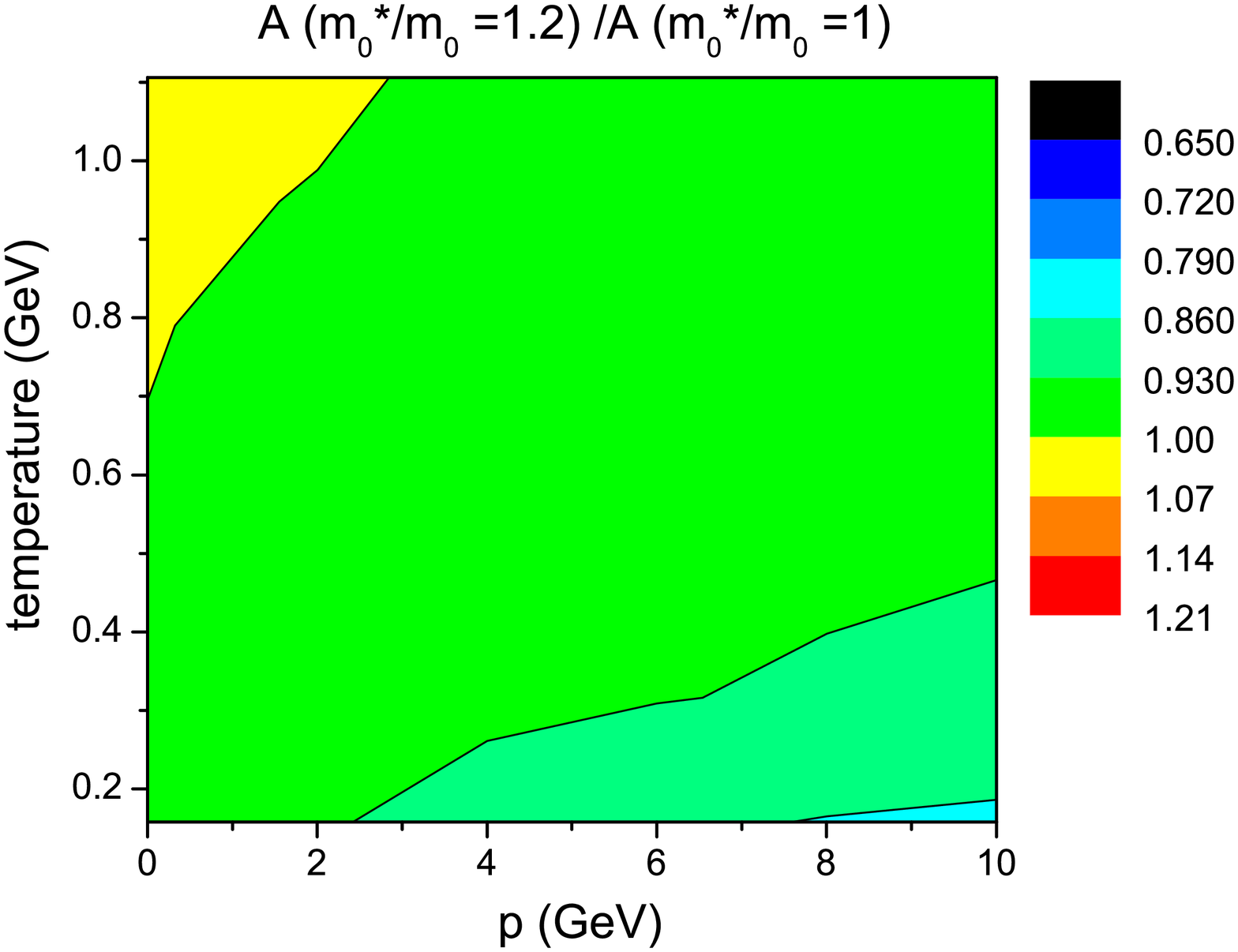}
\includegraphics[width=7.5 cm]{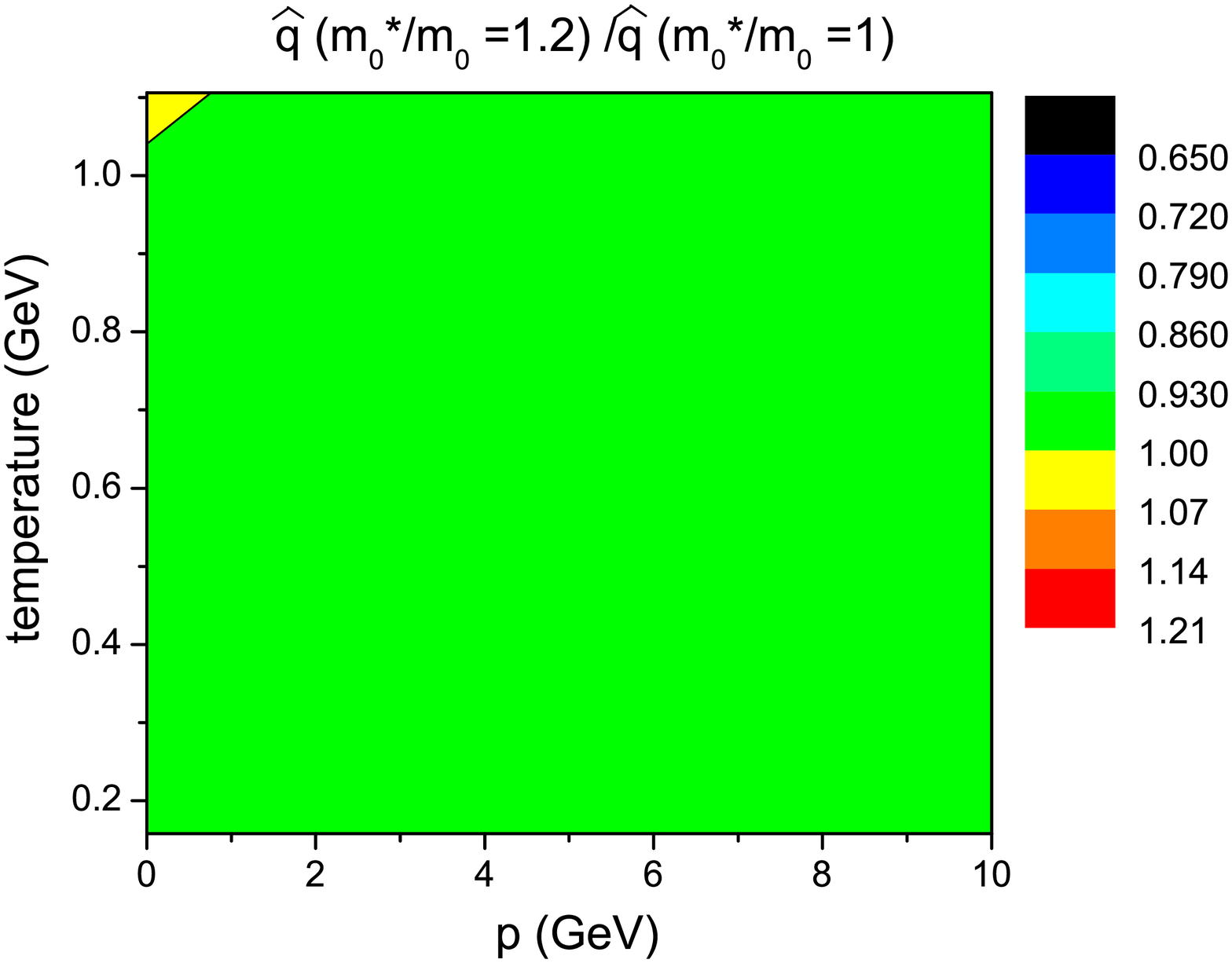}}
\centerline{
\includegraphics[width=7.5 cm]{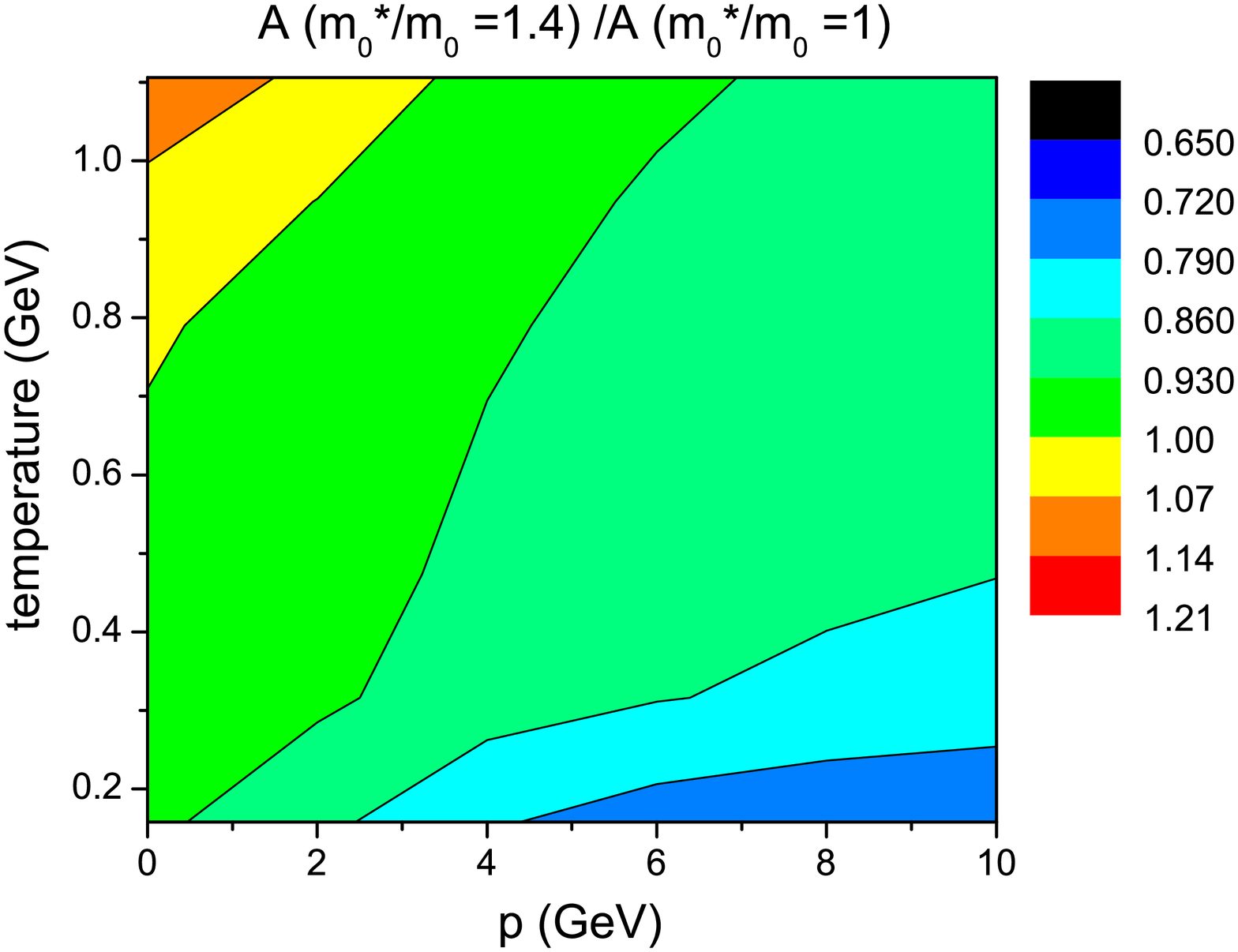}
\includegraphics[width=7.5 cm]{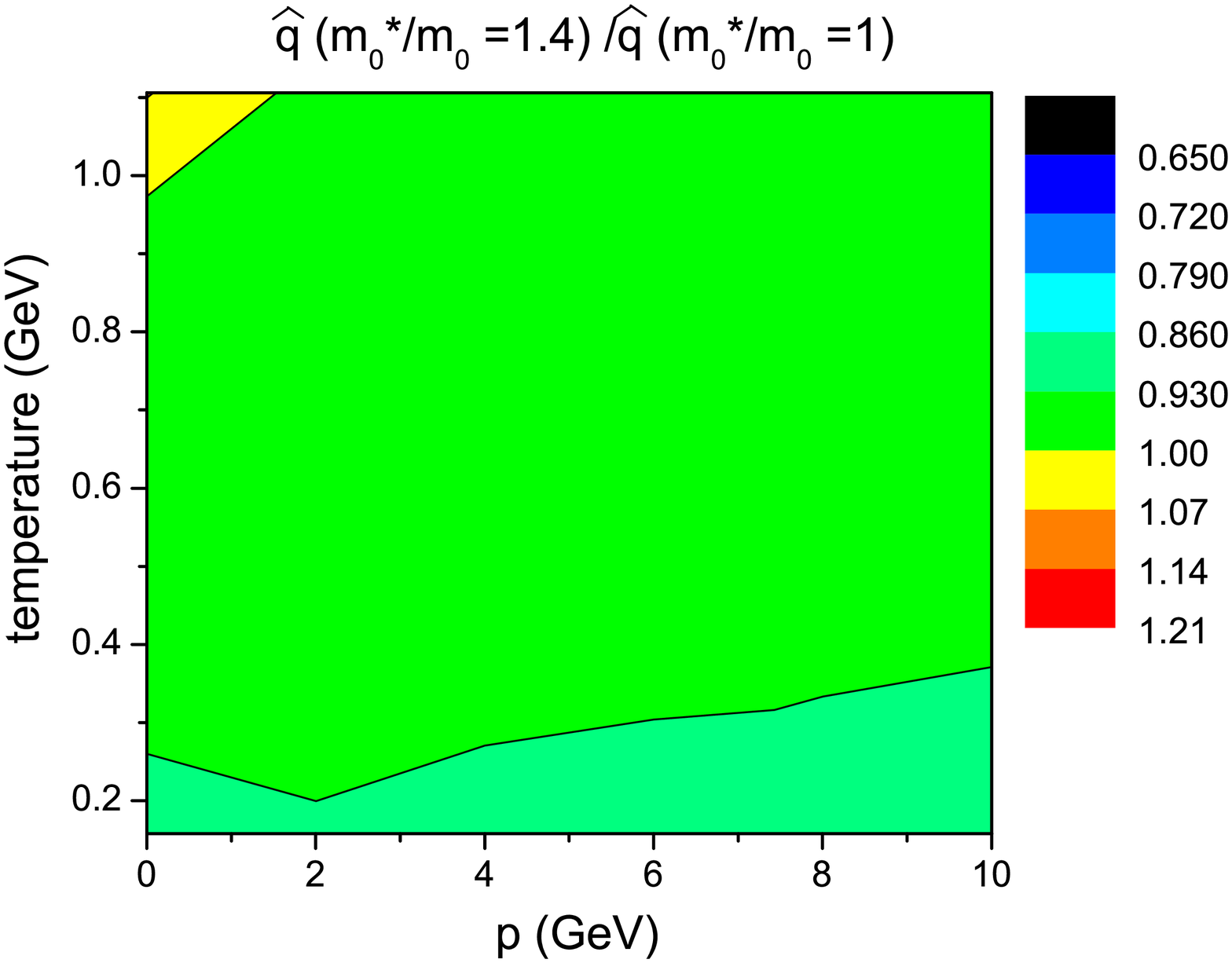}}
\caption{The ratio of the drag coefficient (left) and $\hat{q}$ (right) of a charm quark in a QGP with a pole mass being 0.8, 1.2, and 1.4 times the pole mass calculated from the DQPM and those without shifting the pole mass as a function of the temperature and the charm quark momentum, assuming that the energy density is identical for the modified and for the thermal pole masses.}
\label{ratiom}
\end{figure*}

The results from the DQPM are shown in figure~\ref{ratiom} where the drag coefficient and $\hat{q}$ of charm quarks with pole masses shifted by 0.8, 1.2, and 1.4 times the pole mass from the DQPM are divided by the transport coefficients without shift.
We note that the behavior of the ratios for small charm quark momentum is opposite to that in figure \ref{mass} because of the differences in scattering angle.
The ratio of the drag coefficient is larger than one for a mass of 0.8 times the pole mass and smaller than one for masses of 1.2 and 1.4 times the pole mass.
The deviation from one grows as temperature decreases, partially because the parton number density ratio deviates more from one near $T_c$, as shown in figure~\ref{num-ratiom}.
The results imply that heavy quarks lose more energy in a QGP composed of lighter partons, while they lose less energy with a QGP of heavier partons, assuming that the QGP has the same energy density. The same energy density additionally means that the strong coupling $\alpha_s$ is the same, because in the DQPM it depends only on the temperature (and hence the energy density) and the baryon chemical potential.

\section{summary}\label{summary}

Heavy flavor particles  are one of important probes to investigate the properties of matter under extreme conditions produced in relativistic heavy-ion collisions.
For this we need first of all a proper description of the expansion of the bulk matter. In a second step we can then deduce the transport coefficients of charm quarks by comparing the spectra of charmed mesons with that produced in p+p collisions and by measuring the elliptic flow of the charm quarks.
For describing the expansion of the bulk matter hydrodynamics and transport approaches are most commonly used. The former assumes local thermal equilibrium whereas transport approaches can also handle the situation when the matter is not in equilibrium.

There are two approaches to model the interactions of heavy flavor particles with partonic and hadronic matter. The first is based on the Boltzmann equation which calculates the interaction between bulk particles and heavy quarks microscopically using scattering cross sections of heavy flavor particles. In the second approach it is assumed that this scattering is forward peaked and therefore an expansion in the scattering angle up to second order can be justified. This reduces the Boltzmann equation to a Langevin equation  in which the time evolution of the heavy quark momentum is described by drag and diffusion coefficients in Eqs.~(\ref{A})-(\ref{drag}). Usually these transport coefficients are calculated under the assumption that the expanding matter is always in thermal equilibrium.

In this study we have investigated how the values of these transport coefficients change if the expanding matter is not always in thermal equilibrium.

Our study is based on the dynamical quasi-particle model (DQPM) where QGP is composed of partons which have spectral functions with a pole mass and a width that depend on temperature and baryon chemical potential. The DQPM reproduces the results for lQCD thermodynamics as well as the spatial diffusion coefficient of heavy quarks from  lQCD.

Employing the DQPM, we have studied three non-equilibrium scenarios

i)  We have introduced an anisotropic pressure in the medium by multiplying to parton distribution function a weighting factor which depends on the angle of parton momentum with respect to the beam axis keeping the magnitude of three-momentum constant. This scenario can be parameterized by one parameter $\alpha$, which controls the ratio of the longitudinal and the transverse pressure.  We have found that an anisotropic pressure changes the drag coefficient and $\hat{q}$ of charm quarks in the opposite direction. If the longitudinal pressure is stronger than the transverse pressure, more partons move in longitudinal direction which is perpendicular to charm direction and the charm quark gains longitudinal momentum through scattering, which enhances $\hat{q}$ but suppresses drag coefficient, compared to an isotropic pressure. On the contrary, a stronger transverse pressure than the longitudinal one suppresses $\hat{q}$ and enhances the drag coefficient. Both ratios approach one for a large charm quark momentum. An anisotropic pressure changes the transport coefficients of charm quarks  by at most 20 \%.


ii) We have introduced lower or higher average kinetic energies of the partons as compared to those in thermal equilibrium, keeping the energy density unchanged
by adjusting the parton number density. This scenario is relevant to heavy-ion collisions since at high momenta the particle spectra do not fall off as thermal spectra but have a long tail.
Focusing on charm quarks with large momentum, which is relevant for the study of heavy quark energy loss in the QGP, charm quarks lose more energy per unit length  if colliding with partons of smaller average kinetic energy and hence a larger number density, while the energy loss decreases for partons with larger kinetic energy and hence a smaller number density. The number density is adjusted to have the same energy density in all cases.

The ratio of $\hat q$  shows the opposite trend for charm quarks with a small momentum. The $\hat q$ ratio becomes smaller than one if the kinetic energy of the parton is below the thermal value, and larger than one if it is above. The ratios are largest for small charm momenta where they can deviate from one by 20\%.



iii) We have introduced a non-equilibrium parton mass (in DQPM the parton mass depends on the temperature) by shifting the pole mass of the parton spectral function toward higher or lower masses. The parton number density is adjusted to keep the  energy density constant.
If the pole mass is shifted to lower values, both, drag coefficient and $\hat{q}$ of the heavy quarks, increase, while they decrease if the pole mass is shifted to higher values. 
These ratios deviate from one and the deviation can reach 30 \%  at temperatures close to the critical temperature.
Thus if the matter is composed of lighter partons, heavy quarks suffer a larger energy loss when passing through the medium, while they lose less energy if the medium is composed of heavier partons, assuming that the energy density remains the same.

\begin{figure} [h!]
\centerline{
\includegraphics[width=8.6 cm]{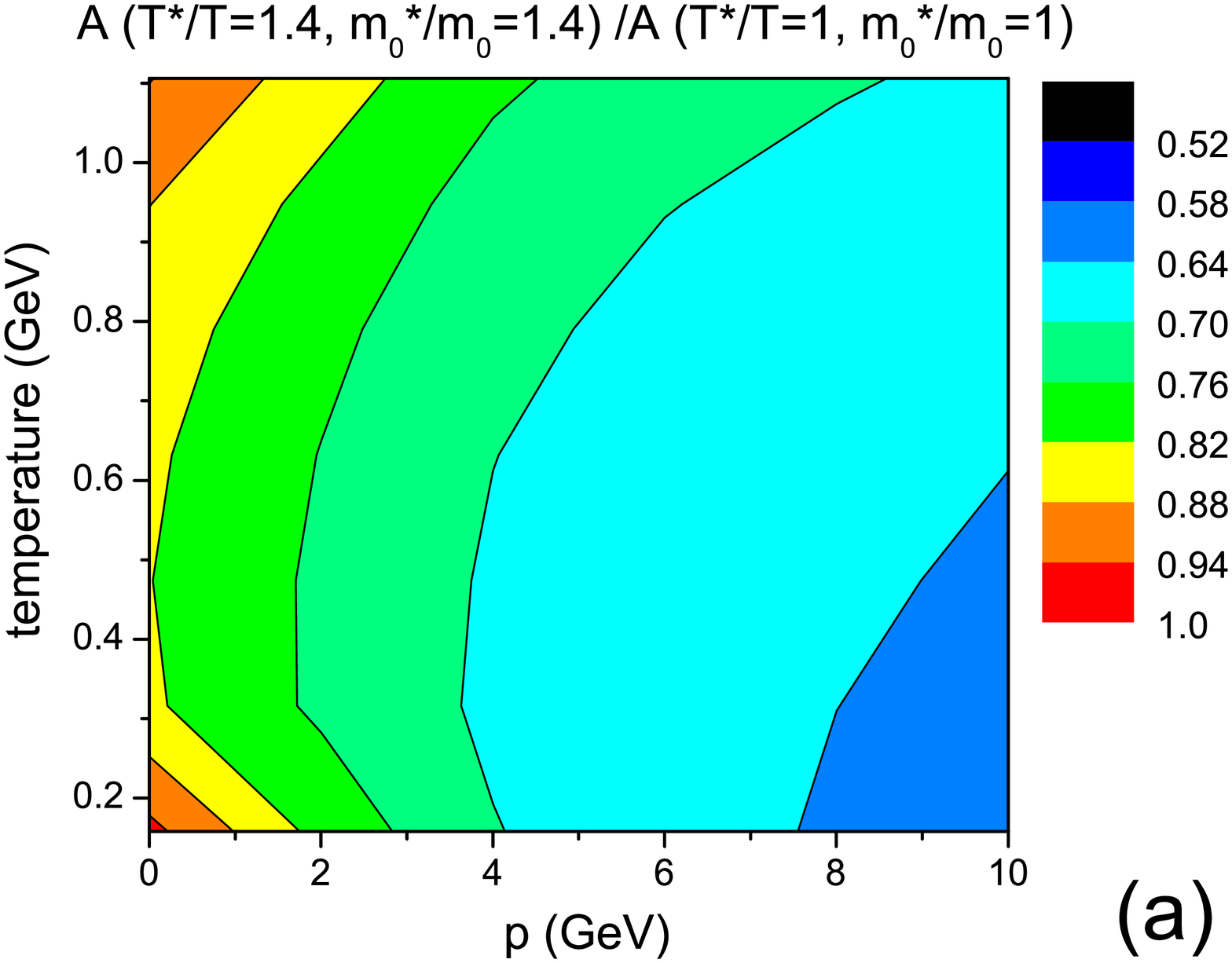}}
\centerline{
\includegraphics[width=8.6 cm]{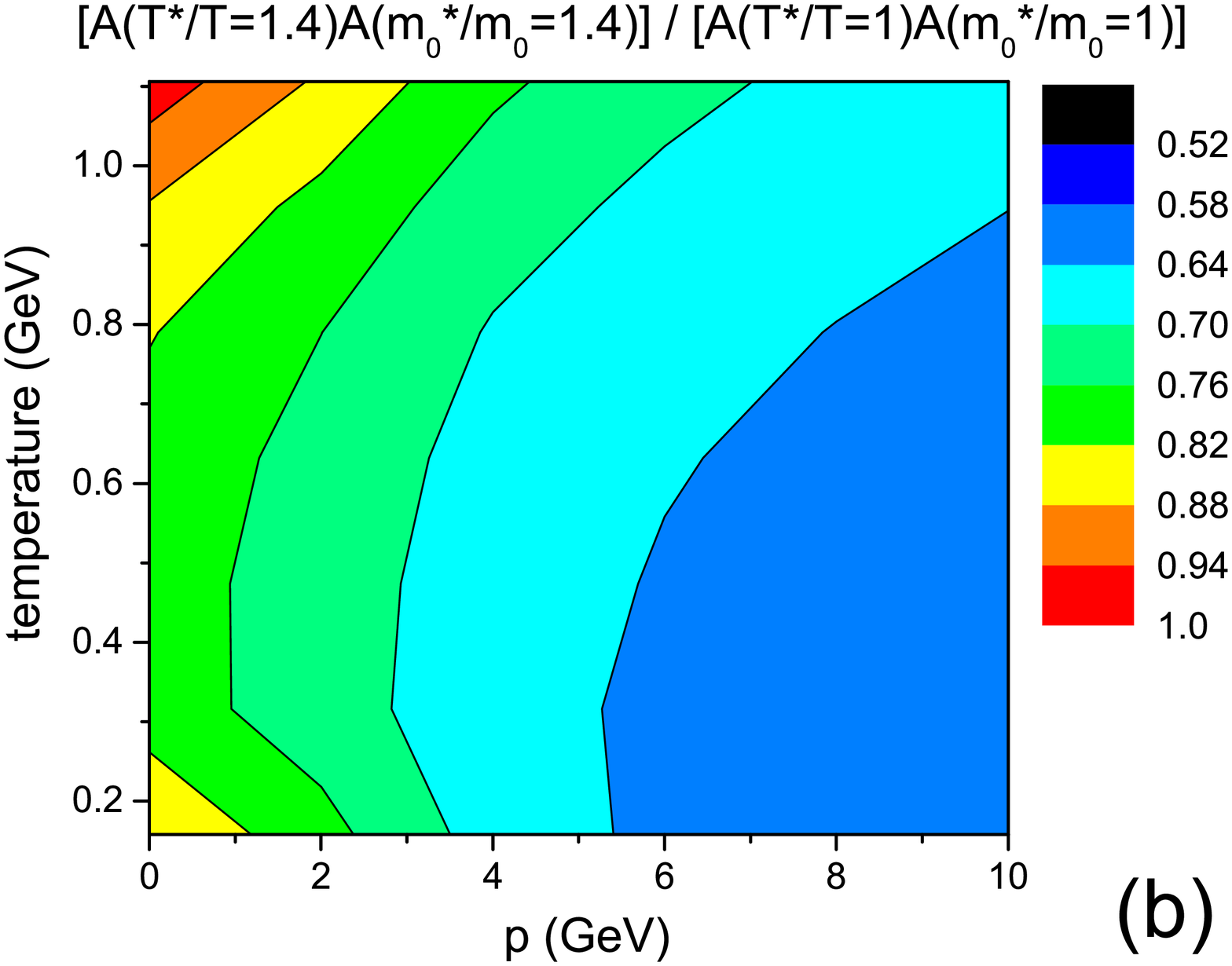}}
\caption{(a) The  drag coefficient of a charm quark in a QGP with the kinetic energy being not in equilibrium with $T^*/T=$ 1.4 and with a pole mass being 1.4 times the pole mass calculated from the DQPM divided by the drag coefficient in equilibrium.  (b) the product  of the ratio of drag coefficients for $T^*/T=$ 1.4 and that of 1.4 times the pole mass.} \label{combination}
\end{figure}

In reality the above three scenarios may coexist in heavy-ion collisions and the effects of one scenario could be constructive or destructive with respect to that from another scenario.
One example to demonstrate this is shown in figure~\ref{combination}.
The upper panel shows the drag coefficient of a charm quark in a QGP with the kinetic energy being not in equilibrium ( $T^*/T=$ 1.4 ) and with a pole mass being 1.4 times the pole mass calculated from the DQPM in equilibrium. The drag coefficient is  divided by its equilibrium value. This corresponds to the combination of the lower left panel of figure~\ref{ratiot} and the lower left panel of figure~\ref{ratiom}. One can see that the drag coefficient of charm quarks is reduced by about 40 \% for large charm quark momenta.
The lower panel of figure~\ref{combination} shows the product of the modification of the drag coefficient  due to a nonequilibrium kinetic energy $T^*/T=$ 1.4 and that due to an increase of the pole mass (taken as 1.4 times the pole mass in equilibrium) . Comparing the upper and lower panel one sees that the change of the drag coefficient for both nonequilibrium effects is almost multiplicative.

Our studies show that the transport coefficients of heavy quarks, calculated by the Boltzmann collision integral, are different for a completely thermalized medium and for a medium
which is only partially equilibrated. For the cases studied in this article the difference is up to 40 \%  and  depends on the momentum of the heavy quark. For an expanding
QGP a complete equilibrium cannot be expected. There are corona particles which cannot come to equilibrium, initially the partons need time to come to equilibrium and the experimental spectra show the presence of high momentum hadrons beyond what is expected from a thermal distribution. Therefore the transport coefficients, obtained by comparing viscous hydrodynamical calculations or Langevin calculations with experimental data, are only effective and may deviate considerably from those one would obtain if the entire QGP
were  in local equilibrium  during the whole expansion.

All these scenarios have in common that the effective transport coefficients, determined by the comparison of experimental data with (hydro)-dynamical calculations, may differ from the true transport coefficient, determined from the cross sections, if the system is not in complete thermal equilibrium. These deviations may reach 30 \%.

Our results have a couple of interesting physical implications.  The first non-equilibrium scenario  is relevant for the anisotropic expansion of matter in relativistic heavy-ion collisions.
The matter, produced around midrapidity when two heavy nuclei pass through each other at extremely high energy, expands quite differently in transverse direction and in beam-direction.
According to ref.~\cite{Florkowski:2013lya}  the longitudinal pressure is  for a while smaller than the transverse pressure. This situation is similar to the lower panels of figure~\ref{anisotropic} ($P_{||}/P_\bot < 1$), where for a large charm quark momentum the drag coefficient of mid-rapidity charm quarks is suppressed as compared to the equilibrium value, while $\hat{q}$ increases.
If one considers that the drag coefficient quantifies the collisional energy loss and that $\hat{q}$ is proportional to the radiative energy loss, an anisotropic pressure lowers the collisional energy loss and enhances the radiative energy loss.

The second and third non-equilibrium scenarios may also be present in heavy ion collisions and affect as well the energy loss of heavy quarks with a large transverse momentum.
The fast expansion of matter in heavy-ion collisions may have the consequence that the pole  mass, which is initially large in the DQPM, cannot adjust itself sufficiently fast to the
equilibrium value. The presence of energetic partons, coming from hard processes, yield higher average kinetic energies than expected in equilibrium. Such a situation corresponds to $T^*/T>1$ and $m^*/m>1$ in figures~\ref{ratiot} and \ref{ratiom}. There both the drag coefficient and $\hat{q}$ of mid-rapidity charm quark decrease at large transverse momentum compared to the equilibrium value. The  consequence is a lower energy loss of  heavy quarks in the QGP, even though the energy density is same.


\section*{Acknowledgements}
The authors acknowledge inspiring discussions with S. Bass, P. B.
Gossiaux, and V. Greco.
This work was supported by the LOEWE center "HIC for FAIR", the HGS-HIRe for FAIR,
the COST Action THOR, CA15213, and the German Academic Exchange Service
(DAAD) (T.S., P.M., E.B.).
Furthermore, PM and EB acknowledge support by DFG through the grant CRC-TR 211 'Strong-interaction matter under extreme conditions'.
The computational resources have been provided by the LOEWE-CSC.
This project has also received funding from the European Union’s Horizon 2020 research and innovation programme under grant agreement No 824093.


\hfil\break
\appendix
\bigskip

\section{}\label{derivations}
In this appendix we derive the interaction rate, drag coefficient, and the $\hat{q}$ of charm quark in the case of Eq.~(\ref{case1}).
For simplicity, scattering amplitude is assumed to be a constant ($|\overline{M}|^2=1$).

\subsection{interaction rate}
If $f(p_2)$  in Eq.~(\ref{int-rate}) is given by
Eq.~(\ref{case1}) , the interaction rate is
\begin{eqnarray}
\frac{dN^{\rm coll}}{dt}= \frac{p^{cm}}{16\pi V E_1E_2\sqrt{s}}.
\end{eqnarray}

Supposing $p_1^\mu=(E_1,~p_1,~0,~0)$ and $p_2^\mu=(E_2,~0,~0,~\pm k)$,

\begin{eqnarray}
s&=&2E_1E_2+m_1^2+m_2^2,\nonumber\\
|\vec{p_1}^{cm}|^2&=&\frac{\{s-(m_1+m_2)^2\}\{s-(m_1-m_2)^2\}}{4s}\nonumber\\
&=&\frac{E_1^2E_2^2-m_1^2m_2^2}{s},
\label{relation1}
\end{eqnarray}
where $E_2=\sqrt{m_2^2+k^2}$.  Expressed in terms of $E_1,~E_2$ the interaction rate is

\begin{eqnarray}
\frac{dN^{\rm coll}}{dt}= \frac{\sqrt{E_1^2E_2^2-m_1^2m_2^2}}{16\pi V E_1E_2 (2E_1E_2+m_1^2+m_2^2)}.
\end{eqnarray}

In the limit of energetic charm partons ($E_1\rightarrow \infty$), the interaction rate is simplified to

\begin{eqnarray}
\lim_{E_1\rightarrow \infty}\frac{dN^{\rm coll}}{dt}= \frac{1}{32\pi V E_1E_2}.
\label{rate-infty}
\end{eqnarray}

Considering the asymptotic form of the scattering cross section

\begin{eqnarray}
\lim_{E_1\rightarrow \infty}\sigma=\lim_{E_1\rightarrow \infty}\frac{1}{16\pi s}=\frac{1}{32\pi E_1E_2},
\label{cs-infty}
\end{eqnarray}
for $|\overline{M}|^2=1$,
Eq.~(\ref{rate-infty}) is the scattering cross section multiplied by particle density, because the relative velocity $v_{ic}\approx 1$ in Eq.~(\ref{def2}).

\subsection{drag coefficient}

We define $p_1^\mu=(E_1,~p_1,~0,~0)$ and $p_2^\mu=(E_2,~0,~0,~k)$. The  boost velocity from the heat bath frame to the center-of-mass frame is given by
\begin{eqnarray}
\vec{\beta}=\bigg(\frac{p_1}{E},~0,~\frac{k}{E}\bigg)
\end{eqnarray}
where $E=E_1+E_2$. The energy-momentum vector of the charm quark in the heat bath frame, $(E_1,~p_1,~0,~0)$, is boosted by $\vec{\beta}$,

\begin{eqnarray}
\widehat{p_1}^{cm}=\frac{1}{|\vec{p_1}^{cm}|}\begin{pmatrix}
-\gamma \beta_x E_1 +\bigg\{1+(\gamma-1)\frac{\beta_x^2}{\beta^2}\bigg\}p_1 \\
0 \\
-\gamma \beta_z E_1 +(\gamma-1)\frac{\beta_x \beta_z}{\beta^2}p_1
\end{pmatrix}.
\label{p1cm}
\end{eqnarray}


Boosting back the vector to the heat bath frame we get
\begin{eqnarray}
\widetilde{L}_i^\mu (\widehat{p_1}^{cm})_\mu=~~~~~~~~~~~~~~~~~~~~~~~~~~~~~~~~~~~~~\nonumber\\
\nonumber\\
\begin{pmatrix}
\gamma\beta_x(\widehat{p_1}^{cm})_x+\gamma\beta_z(\widehat{p_1}^{cm})_z\\
\bigg\{1+(\gamma-1)\frac{\beta_x^2}{\beta^2}\bigg\}(\widehat{p_1}^{cm})_x+(\gamma-1)\frac{\beta_x\beta_z}{\beta^2}(\widehat{p_1}^{cm})_z \\
0\\
(\gamma-1)\frac{\beta_x\beta_z}{\beta^2}(\widehat{p_1}^{cm})_x+\bigg\{1+(\gamma-1)\frac{\beta_z^2}{\beta^2}\bigg\}(\widehat{p_1}^{cm})_z
\end{pmatrix},\nonumber\\\label{middle}
\\
\nonumber
\end{eqnarray}
with $(\widehat{p_1}^{cm})_\mu=(0,~\vec{\widehat{p_1}}^{cm})$.
Since $(\widehat{p_1})_i=\widehat{x}$ in Eq.~(\ref{final-A}), only $x$-component contributes, and the $z$-component is cancelled by the second term in Eq.~(\ref{case1}).
Substituting Eq.~(\ref{p1cm}) into Eq.~(\ref{middle}), and then Eq.~(\ref{middle}) into Eq.~({\ref{final-A}), and using Eq.~(\ref{relation1}) and the relation

\begin{eqnarray}
\gamma=\frac{E_1+E_2}{\sqrt{2E_1E_2+m_1^2+m_2^2}}.
\label{relation2}
\end{eqnarray}
The drag coefficient is given by

\begin{eqnarray}
A=\frac{(E_1E_2+m_2^2)\sqrt{E_1^2E_2^2-m_1^2m_2^2}}{16\pi V E_1E_2(2E_1E_2+m_1^2+m_2^2)^2}~p_1.
\label{drag3}
\end{eqnarray}

For an extremely energetic charm quark, the drag coefficient is

\begin{eqnarray}
\lim_{E_1\rightarrow \infty}A=\frac{1}{64\pi V E_2},
\label{drag-infty}
\end{eqnarray}
which is the product of the interaction rate of Eq.~(\ref{rate-infty}) and

\begin{eqnarray}
\lim_{E_1\rightarrow \infty}
\widetilde{L}_x^\mu (\widehat{p_1}^{cm})_\mu p_{cm}=\frac{p_1}{2}\approx \frac{E_1}{2}
\label{drag-boost}
\end{eqnarray}
from Eqs.~(\ref{dpl}) and (\ref{middle}).

\subsection{$\hat{q}$}

As in the calculations of the drag coefficient, boosting back to the heat bath frame from the center-of-mass frame, the transverse component of the transport coefficient is expressed by Eqs.~(\ref{tensor}) and (\ref{general}) as

\begin{eqnarray}
&&\bigg\{\delta_{ij}-(\widehat{p_1})_i(\widehat{p_1})_j\bigg\}\widetilde{L}_{i}^{\mu}\widetilde{L}_{j}^{\nu} \langle\Delta p_\mu^{cm}\Delta p_\nu^{cm}\rangle,\nonumber\\
&&=\bigg\{\delta_{ij}-(\widehat{p_1})_i(\widehat{p_1})_j\bigg\}\nonumber\\
&&\times\widetilde{L}_{i}^{\mu}\widetilde{L}_{j}^{\nu}\bigg\{\frac{\delta_{\mu\nu}}{3}+(\widehat{p_1}^{cm})_\mu(\widehat{p_1}^{cm})_\nu\bigg\}(p^{cm})^2,
\label{qhat1}
\end{eqnarray}
where $\delta_{0\mu}=\delta_{\mu 0}=0$ and $(\widehat{p_1}^{cm})_0=0$.
Since the charm quark moves in $x$-direction, Eq.~(\ref{qhat1}) gives

\begin{eqnarray}
\frac{(p^{cm})^2}{3}\sum_{k=x,y,z}(\widetilde{L}_{y}^{k}\widetilde{L}_{y}^{k}+\widetilde{L}_{z}^{k}\widetilde{L}_{z}^{k})+ \bigg\{p^{cm}\widetilde{L}_{z}^{\mu}(\widehat{p_1}^{cm})_\mu\bigg\}^2\nonumber\\
=\frac{(p^{cm})^2}{3}(2+\gamma^2\beta_z^2)+\bigg\{\gamma^2\beta_z^2(-E_1+\beta_x p_1)\bigg\}^2,
\end{eqnarray}
where $\widetilde{L}_{y}^{\mu}(\widehat{p_1}^{cm})_\mu=0$ as shown in Eq.~(\ref{middle}).
Therefore, $\hat{q}$ is given by

\begin{eqnarray}
\hat{q}=\frac{(p^{cm})^3}{16\pi V p_1E_2\sqrt{s}}~~~~~~~~~~~~~~~~~~~~~~~~~~~~~~~~~~~~~~~~~\nonumber\\
\times \bigg\{\frac{2+\gamma^2\beta_z^2}{3}+\frac{\gamma^4\beta_z^4}{(p^{cm})^2}(-E_1+\beta_x p_1)^2\bigg\}~~~~~~~~~~~\nonumber\\
=\frac{k^2(E_1^2E_2^2-m_1^2m_2^2)^{3/2}}{48\pi V p_1E_2(2E_1E_2+m_1^2+m_2^2)^3}~~~~~~~~~~~~~~~~~\nonumber\\
\times\bigg\{\frac{2(2E_1E_2+m_1^2+m_2^2)}{k^2}+1+\frac{3(E_1E_2+m_1^2)^2}{E_1^2E_2^2-m_1^2m_2^2}\bigg\}.\nonumber\\
\end{eqnarray}

Using Eqs.~(\ref{relation1}) and (\ref{relation2}), $\hat{q}$ of an energetic charm quark is approximately given by

\begin{eqnarray}
\lim_{E_1\rightarrow \infty}\hat{q}=\frac{1}{96\pi V},
\label{qhat-infty}
\end{eqnarray}
which is the interaction rate of Eq.~(\ref{rate-infty}) multiplied by

\begin{eqnarray}
\lim_{E_1\rightarrow \infty}\langle(\Delta p_T^{cm})^2\rangle=\lim_{E_1\rightarrow \infty}\frac{2}{3}(p^{cm})^2=\frac{E_1E_2}{3}
\label{dpt2b}
\end{eqnarray}
from Eq.~(\ref{dpt2}). We note that $\langle(\Delta p_T^{cm})^2\rangle$ does not need a Lorentz boost, different from the calculations of drag coefficient in Eq.~(\ref{drag-boost}).

\end{document}